\documentclass[12pt]{JHEP3}

\usepackage{epsfig,amsfonts,amssymb,amsmath,color}

 \usepackage{marginnote}
 \reversemarginpar
 \newcommand{\newnote}{ \marginnote{\bf New in v2.0:}[-4mm]}
 
\newcommand{\mathematica}[3]{\vspace{0.35cm}\noindent\boxed{\begin{minipage}{#1\textwidth}\begin{tabular}{lp{13cm}}{\color{paper_blue}{\scriptsize{\tt In[1]:}}\raisebox{-0.65pt}{{\scriptsize{\tt=}}}}&{\tt #2}\\{\color{paper_blue}{\scriptsize {\tt Out[1]:}}\raisebox{-0.65pt}{{\scriptsize{\tt=}}}}&{\tt #3}\end{tabular}\end{minipage}}\vspace{0.35cm}}

\definecolor{varcolor}{rgb}{0.1,0.55,0.25}
\definecolor{functioncolor}{rgb}{0.1,0.35,0.75}
\definecolor{paper_blue}{rgb}{0.3,0.2,0.75}
\definecolor{paper_red}{rgb}{0.65,0.1,0.15}
\definecolor{paper_green}{rgb}{0.05,0.35,0.125}
\definecolor{paper_grey}{gray}{0.375}
\definecolor{perm}{rgb}{0.1,0.45,0.85}
\definecolor{deemph}{rgb}{0.7,0.7,0.7}
\newcommand{\vardef}[1]{{\color{varcolor}{\sl #1}\rule[-1.05pt]{7.5pt}{.75pt}}}

\newcommand{\defn}[3]{~\\[-35pt]\begin{itemize}\item[]\indent\hspace{-21pt}$\bullet$\hspace{-.75pt} {\tt {\color{functioncolor}#1}\![}#2{\tt\,]\!:}#3\end{itemize}\vspace{-7pt}
}

\newcommand{\defvar}[2]{~\\[-30pt]\begin{itemize}\item[]\indent\hspace{-21pt}$\bullet$\hspace{-.75pt} \var{#1}: #2\end{itemize}\vspace{-10pt}}
\newcommand{\var}[1]{{\tt{\color{varcolor}{\sl#1}}}}

\newcommand{\fun}[1]{{\color{functioncolor}#1}}

\newcommand{\bZ}{\mathbb{Z}}
\newcommand{\cC}{\mathcal{C}}
\newcommand{\cQ}{\mathcal{Q}}
\newcommand{\IZ}{\mathbb{Z}}
\newcommand{\kk}{k}
\newcommand{\IR}{\mathbb{R}}

\newcommand{\be}{\begin{equation}}
\newcommand{\ee}{\end{equation}}
\newcommand{\ben}{\begin{eqnarray}\displaystyle}
\newcommand{\een}{\end{eqnarray}}

\newcommand{\refb}[1]{(\ref{#1})}
\newcommand{\p}{\partial}
\newcommand{\sectiono}[1]{\section{#1}\setcounter{equation}{0}}

\newcommand{\NN}{{\cal N}}
\newcommand{\cN}{\mathcal{N}}
\newcommand{\cM}{\mathcal{M}}
\def\Tr{\,{\rm Tr}\, }

\newcommand{\wh}{\widehat}

\newcommand{\ta}{\tilde\alpha}
\newcommand{\tc}{\tilde d}
\newcommand{\ha}{\hat\alpha}
\newcommand{\ca}{\check \alpha}
\newcommand{\cc}{\tilde c}
\newcommand{\dx}{c}

\newcommand{\gref}{g_{\rm Coulomb}}
\newcommand{\gR}{G_{\rm Higgs}}
\newcommand{\gRa}{g_{\rm Higgs}}
\newcommand{\QR}{Q_{\rm Higgs}}
\newcommand{\bQR}{\bar Q_{\rm Higgs}}
\newcommand{\QC}{Q_{\rm Coulomb}}
\newcommand{\bQC}{{\bar Q}_{\rm Coulomb}}
\newcommand{\gC}{G_{\rm Coulomb}}
\newcommand{\OmS}{\Omega_{\rm S}}  

\def\newdefy{\vardef{y}}

\title{On the Coulomb and Higgs branch formulae for multi-centered black holes
and quiver invariants}
 
\preprint{  
Bonn-TH-2013-03\\
CERN-PH-TH/2013-030\\
HRI/ST/1302\\
arXiv:1302.5498v2}

\author{Jan Manschot$^{1,2}$, Boris Pioline$^{3,4}$, Ashoke Sen$^{5}$
\\
$^1$ {\it Bethe Center for Theoretical Physics, Universit\"at Bonn, Nu\ss allee 12 \\
53115 Bonn, Germany}\\

$^2$ {\it Max Planck Institute for Mathematics, Vivatsgasse 7, 53111 Bonn, Germany} \\

$^3$ {\it CERN PH-TH,
Case C01600, CERN, CH-1211 Geneva 23, Switzerland}\\

$^4$ {\it Laboratoire de Physique Th\'eorique et Hautes
Energies, CNRS UMR 7589, \\
Universit\'e Pierre et Marie Curie,
4 place Jussieu, 75252 Paris cedex 05, France} \\

$^5$ Harish-Chandra Research Institute,
Chhatnag
Road, Jhusi, Allahabad 211019, India
\\

\vspace*{2mm} {\tt e-mail: \email{
manschot@uni-bonn.de, boris.pioline@cern.ch,
sen@hri.res.in}
} \vspace*{-3mm}

}

\abstract{
In previous work we have shown that the equivariant index of 
multi-centered $\cN=2$ black holes localizes on collinear configurations 
along a fixed axis. Here we provide a general algorithm for enumerating such 
collinear configurations and computing their contribution to the index.
We apply this machinery to the case of black holes described by quiver 
quantum mechanics, and give a systematic prescription -- the Coulomb branch
formula -- for computing the  
cohomology of the moduli space of quiver representations. For quivers 
without  oriented loops,  the Coulomb 
branch formula is shown to agree with the Higgs branch formula based on Reineke's result for 
stack invariants,
even when the dimension vector is not primitive. For quivers with  oriented loops,
the Coulomb branch formula parametrizes the Poincar\'e polynomial of the quiver moduli space in 
terms of single-centered (or pure-Higgs) BPS invariants, which are conjecturally
independent of the stability condition (i.e. the choice of Fayet-Iliopoulos parameters) 
and angular-momentum free. To facilitate further investigation  
we provide a {\sc Mathematica} package ``{\tt CoulombHiggs.m}'' implementing 
the Coulomb and Higgs branch formulae.
}

\begin{document}

\tableofcontents

\sectiono{Introduction} \label{s1}

BPS states in $\NN=2$ supersymmetric string vacua offer a rich playground 
for exploring the
microscopic properties of black holes in quantum gravity. Indeed, such states have
dual descriptions as black hole solutions in supergravity at strong coupling, and brane
configurations at weak coupling. The Witten index 
(more precisely, the second helicity supertrace\cite{Bachas:1996bp,Gregori:1997hi}) 
remains unchanged upon varying the string coupling. 
Computing the index $\Omega(\gamma)$ in both regimes as function of the
electromagnetic charges $\gamma$  
can provide non-trivial tests of the 
equivalence between the microscopic and gravitational 
descriptions expected in any theory of quantum gravity \cite{Strominger:1996sh}. 

This comparison however is  
complicated by the fact that on the macroscopic side,
contributions to the index $\Omega(\gamma)$ originate not only from single-centered
black holes with charge $\gamma$, but also from multi-centered black holes 
with constituents carrying charges $\{\alpha_i\}$  such that  
$\gamma=\sum \alpha_i$ \cite{0005049,0206072,0304094,0702146}. 
The total  index carried by single and multi-centered black holes 
is the quantity that should be compared with the index on the 
microscopic side. In contrast, the gravitational path integral in the  near horizon
geometry of  each black hole gives information only about the index 
$\OmS(\alpha)$ associated with single-centered 
black holes\cite{0809.3304,Sen:2009vz}. Thus it is necessary  
to express the 
contribution from multi-centered black holes
in terms of the index $\OmS(\alpha_i)$ associated
to each center. For this 
it suffices to compute the index\footnote{The Fayet-Iliopoulos (FI) parameters $\{c_i\}$ depend 
on  $\alpha_i$ and on the values of the moduli at infinity, and will be treated as 
independent real parameters subject to the constraint $\sum_i c_i=0$.} 
$\gref(\{\alpha_i\},\{c_i\})$
of the supersymmetric quantum mechanics of $n$ centers 
carrying charges $\{\alpha_i\}$ 
interacting by Coulomb and Lorentz forces (and other forces related by supersymmetry). 
This `Coulomb index'   
was computed   in \cite{1011.1258,1103.1887},
leading to a general prescription -- the Coulomb branch formula -- for expressing 
the total index $\Omega(\gamma)$ in terms of the single-centered 
BPS invariants $\OmS(\alpha_i)$. 
Unlike
these single-centered BPS invariants, the Coulomb index
$\gref(\{\alpha_i\},\{c_i\})$ depends on the moduli at infinity -- 
indeed, this dependence is responsible for jumps of the total index $\Omega(\gamma)$ 
across walls of marginal 
stability, providing a physically transparent derivation of the wall-crossing formulae known in the mathematical literature \cite{0811.2435,1011.1258,Pioline:2011gf}. 

An important technical tool
in \cite{1011.1258,1103.1887} was to consider a `refined' version
of these indices, which computes the trace $\Tr'(-1)^F y^{2J_3}$,
where $J_3$ is the angular momentum generator with respect to a fixed $z$-axis,
and the prime denotes the removal of 
fermion zero mode contributions before taking the trace.
Unlike the helicity supertrace 
$\Tr'(-1)^F$, the refined index is not 
a protected quantity in full-fledged string theory and cannot 
be directly compared with the microscopic results. This generalization was nonetheless
necessary, as one could use localization  with respect to rotations along the $z$ axis
to compute the refined Coulomb index 
$\gref(\{\alpha_i\},\{c_i\};y)$ for any number of centers $n$, and then take the limit $y\to 1$ 
at the end to recover the result for $\Tr'(-1)^F$. 
Although the description of the fixed points of this action is straightforward, 
their enumeration requires solving a set of $n-1$ algebraic equations in $n-1$ real
variables, which quickly becomes unpractical as the number of centers increases. The 
first goal of this paper is to remove this bottleneck and give a completely algorithmic way of  computing the Coulomb index $\gref(\{\alpha_i\};\{c_i\};y)$, 
and hence the total refined index 
$\Omega(\gamma;y)$.\footnote{In \cite{1011.1258,1103.1887,Manschot:2012rx} 
we used a subscript $~_{\rm ref}$ to denote a refined index. In this paper we shall drop
this subscript to avoid cluttering up the formul\ae, but it should be understood that
whenever the index carries the argument $y$, it corresponds to a refined index.}
Using the connection between multi-centered black hole quantum mechanics and
quiver quantum mechanics described in \cite{Manschot:2012rx} this leads to an explicit
expression for the Poincar\'e polynomial of quiver moduli spaces. 
The second 
goal is to establish the equivalence of this Coulomb branch
formula with 
Reineke's formula\cite{MR1974891} and its generalizations\cite{Joyce:2004,0811.2435}
for the Poincar\'e polynomial of quiver
moduli spaces for general quivers with no oriented  loop.

\subsection*{Coulomb index from localization}

Before explaining our new algorithm, let us briefly review the prescription of  \cite{1103.1887}  for computing the refined Coulomb index $\gref(\{\alpha_i\},\{c_i\};y)$. 
Let $\Gamma$ be the charge  lattice of electromagnetic charges, 
equipped with  the Dirac-Schwinger-Zwanziger (DSZ) symplectic product $\langle \cdot,\cdot
\rangle \in \IZ$.
Consider a multi-centered black hole configuration where each center carries
charge $\alpha_i$  ($i=1\dots n$), with DSZ products
$\alpha_{ij}\equiv \langle \alpha_i,\alpha_j\rangle$ between the charges. 
For fixed values of the moduli at infinity, encoded in the FI parameters $c_i$, 
$n$-centered configurations are parametrized by a 
$2n-2$-dimensional phase space
$\cM_n(\{\alpha_{ij}\},\{\dx_i\})$  \cite{0005049,0206072}. 
The coordinates of $\cM_n$ are  
the locations of  $n$ centers 
$\vec r_i$ up to overall translations,
subject to $n-1$ constraints\footnote{The constraint 
\eqref{eor0} for $i=n$ follows from 
the sum of others using the fact that  $\sum_{i=1}^n \dx_i=0$. For
a general multi-centered black hole system there are additional constraints 
besides \refb{eor0} coming
from the requirement of the
regularity of the metric and other fields. However when the central charges
of the constituents nearly align, which is the the limit in which
the quiver quantum mechanics becomes a good description, these additional 
constraints are expected to be satisfied automatically 
\cite{Manschot:2012rx}. Throughout this paper we shall be working in this limit.}
\be
\label{eor0}
\forall i=1\dots n\ ,\quad \sum_{j \atop  
j\neq i} \frac{\alpha_{ij}}{|\vec r_i -\vec r_j|} = \dx_i\ .
\ee
The space $\cM_n$ admits a symplectic form \cite{0206072,0807.4556,0906.0011}, 
such that  the action
of $SO(3)$ rotations on $\cM_n$ is generated by the moment map
\be
\vec J = \frac12 \sum_{i<j} \alpha_{ij} \frac{\vec r_j -\vec r_i}{|\vec r_i -\vec r_j|} \ ,
\ee
equal to the angular momentum carried by the configuration. 
We denote by
\be
\label{defgref}
\gref(\{\alpha_1,\cdots \alpha_n\};\{\dx_1,\cdots \dx_n\};y)=\Tr'(-y)^{2J_3}
\ee
the refined index 
associated to this multi-centered configuration 
{\it assuming that all the centers are distinguishible from each other, and
that each center carries no intrinsic 
degeneracy}.
Mathematically, this is  
the equivariant Dirac index of the symplectic space 
$\cM_n(\{\alpha_{ij}\},\{\dx_i\})$ \cite{1103.1887,1107.0723}. 
We refer to \eqref{defgref} 
as the Coulomb index of multi-centered black holes with charges  $\{\alpha_i\}$.
This in turn can be used to compute the 
index associated with a general multi-centered black hole configurations in terms
of indices $\OmS(\alpha_i)$ carried by individual centers following the  
procedure described in \cite{1103.1887,Manschot:2012rx}.

When there exists an ordering of the charges $\{\alpha_i\}$ such that $i\leq j$ 
if and only if $\alpha_{ij}\ge 0$, and for generic values of the parameters $c_i$ 
away from walls of marginal stability, 
the symplectic space $\cM_n$ 
is compact, and
the Coulomb index can be computed by localization with respect to rotations around a fixed axis, 
using the Atiyah-Bott Lefschetz fixed point formula \cite{1011.1258,1103.1887}.
The configurations which stay invariant under such rotations are collinear configurations,
corresponding to solutions of \eqref{eor0} lying along the $z$ axis.
The result is expressed as 
\be \label{eindex1}
\gref(\{\alpha_1,\cdots \alpha_n\}; \{\dx_1,\cdots \dx_n\};y) 
= (-1)^{n-1+\sum_{i<j} \alpha_{ij} } (y-y^{-1})^{-n+1} \sum_{\rm extrema}\, \pm 
\, y^{\sum_{i<j} 
\alpha_{ij} \, {\rm sign}(z_j-z_i)} \, ,
\ee
where the sum runs over solutions to the equations
\be \label{eor1}
\sum_{j=1\atop j\ne i}^n {\alpha_{ij}\over |z_i - z_j|}  =  \dx_i\, , \quad \hbox{for $1\le i\le n-1$}\, \ ,
\quad z_1=0\ .
\ee
The $z_1=0$ condition fixes the translational zero-mode.
The sign $\pm$ in \eqref{eindex1} is given by the sign of the Hessian 
$\det (\p^2 W / \p z_i \p z_j)$  
of the superpotential
\be \label{ecou0}
W(\{z_i\})=-\sum_{i<j} \alpha_{ij} \, {\rm sign}(z_j-z_i) \log| z_i- z_j|
-\sum_{i=1}^n \dx_i\, z_i \, ,
\ee
whose critical points reproduce the conditions \eqref{eor1}.

For charges $\{\alpha_i\}$ such that no such ordering exists, the space  $\cM_n$
may be non-compact, due to the possibility of `scaling solutions', i.e. a subset of the centers 
approaching each other at arbitrary short distances \cite{Bena:2006kb,0702146}. In that case,
we continue to define  the Coulomb index $\gref(\{\alpha_1,\cdots \alpha_n\}; \{\dx_1,\cdots \dx_n\};y)$ 
by the localization formula \eqref{eindex1}, although the result can no longer be 
interpreted directly as the refined index 
associated to the multi-centered black hole configuration (in particular, it  may not be 
a symmetric Laurent polynomial). Nevertheless it can be used
to construct such a refined index following the procedure described in
\cite{1103.1887,Manschot:2012rx} and reviewed in \S\ref{sec_Poincare}.
When the FI parameters $c_i$ sit on a wall of marginal stability, the space $\cM_n$
is also non-compact due to the possibility of separating the centers into two or more 
clusters at arbitrarily large distances, and we do not assign a value to 
$\gref(\{\alpha_1,\cdots \alpha_n\}; \{\dx_1,\cdots \dx_n\};y)$ in such cases.

\subsection*{A new algorithm for computing the Coulomb index}

Except in very special cases, it is usually impossible to find all solutions of \eqref{eor1} explicitly.
This is also unnecessary, since the contribution of a given solution of  \eqref{eor1} to the total
index \eqref{eindex1} depends only on the ordering of the centers, via the angular momentum
$J_3=\frac12 \sum_{i<j}\, \alpha_{ij}\, 
{\rm sign}(z_j-z_i)$ along the $z$-axis and the sign of the Hessian 
$W''$. For a small number of centers, it is possible to find approximate solutions numerically, 
and determine both $J_3$ and the sign of $W''$, however this becomes quickly unpractical as the number of centers grows. 
Moreover, the brute force enumeration of solutions of \eqref{eor1} does not take into account the fact that there can be cancellations between different solutions with the same ordering. To exploit this fact, it is useful to  associate a permutation $\sigma$ to each solution to \refb{eor1}, such that $i<j$ iff $z_{\sigma(i)}<z_{\sigma(j)}$. Defining $\ta_i =\alpha_{\sigma(i)}$, $x_i=z_{\sigma(i)}$ and $\cc_i=\dx_{\sigma(i)}$,  solutions of \eqref{eor1}
correspond to critical points of 
\be
\label{ecou1}
W(\{x_i\})=-\sum_{i<j} \ta_{ij} \log| x_i- x_j|
-\sum_{i=1}^n \cc_i\, x_i \, , \quad \sum_{i=1}^n \cc_i=0\
\ee
in the physical region 
\be
0 \equiv x_1<x_2<x_3\cdots < x_n\, .
\ee
Reorganizing \eqref{eindex1} as a sum over all permutations 
$\sigma$ of $1,2,\dots n$, we obtain
\be \label{eindex}
\gref(\{\alpha_1,\cdots \alpha_n\}; \{\dx_1,\cdots \dx_n\};y) = (-1)^{n-1+\sum_{i<j} \alpha_{ij}} (y-y^{-1})^{-n+1} \sum_\sigma s(\sigma)\, 
y^{\sum_{i<j} 
\alpha_{\sigma(i)\sigma(j)}} \, ,
\ee
where $s(\sigma)$ is the sum of the sign of the Hessian of \eqref{ecou1}
over each critical point in the physical region. In particular, $s(\sigma)$ is 
insensitive to pairs of solutions of \eqref{eor1} with the same ordering,
which may appear under small perturbations of the parameters $\alpha_{ij}$ and $\dx_i$,
as long as we stay away from walls of marginal stability in the space of 
FI parameters $\{c_i\}$ and away from certain `scaling walls' in the space of DSZ products
$\{\alpha_{ij}\}$ described in more detail below.

The first aim of this paper is to develop an explicit algorithm 
for computing $s(\sigma)$ 
and hence the Coulomb index \eqref{eindex} for generic DSZ products  $\{\alpha_{ij}\}$ and FI parameters $\{\dx_i\}$.
This is achieved in \S\ref{sec_Coulomb}, where
we prove an inductive formula \eqref{efinal1} for the indexed number 
$s(\sigma)=F(\{\ta_1,\cdots \ta_n\}, \{\cc_1,\cdots \cc_n\})$ of critical points of the superpotential
\eqref{ecou1}, by exploiting its robustness under changes of the DSZ products $\ta_{ij}$.
For quivers without oriented loops and generic products, we arrive at the following completely
explicit result:
\be \label{eindexfinint}
\begin{split}
 \gref(\{\alpha_1,\cdots \alpha_n\}; \{\dx_1,\cdots \dx_n\};y) 
=& (-1)^{n-1+\sum_{i<j} \alpha_{ij}} (y-y^{-1})^{-n+1} \\
 \sum_\sigma 
\prod_{k=1}^{n-1}  \Theta\left(\alpha_{\sigma(k), \sigma(k+1)} \,  \sum_{i=1}^k c_{\sigma(i)} \right) &
(-1)^{\sum_{k=1}^{n-1}
\Theta(-\alpha_{\sigma(k),\sigma(k+1)})}\, 
y^{\sum_{i<j} 
\alpha_{\sigma(i)\sigma(j)}} \, ,
\end{split}
\ee
where $\Theta$ is the Heaviside function and the sum runs over all permutations
$\sigma$ of $1,2,\cdots n$. 
If some of the $\alpha_{ij}$'s vanish
then we need to deform them away from zero such that the deformed quiver continues
to satisfy the no loop condition, compute the result using \refb{eindexfinint} and then take
the limit where the deformations are taken back to zero.

As described in \refb{efinal1}, 
in the presence of loops the expression for $s(\sigma)$ 
picks up additional contributions $\Delta F_A$ 
given  
in \eqref{edeltafs1}, 
which depend on the index $F$ with
fewer centers and another auxiliary quantity $G(\ta_1,\cdots \ta_n)$. The latter
counts the (indexed) number\footnote{Due to the scaling symmetry, each solution arises as a
one-parameter family, the number of which is counted by $G$.} of collinear scaling solutions,  
i.e. solutions of \eqref{eor1} with $\dx_i=0$ which may arise when the total angular momentum $\frac12\sum_{i<j} \ta_{ij}$ vanishes. We find that this
`scaling index' can itself be computed inductively using \eqref{egrecur}. 
These formul\ae\ hold when the  DSZ products $\ta_{ij}$ are generic, but we show that even
when this is not the case, 
all the relevant physical quantities can be computed in terms of limits of
these formul\ae.

\subsection*{Quiver quantum mechanics and pure-Higgs invariants}

While the refined index is not protected in full-fledged string theory, it is protected in the context of
$\NN=4$ supersymmetric quiver quantum mechanics, which
describes the dynamics of certain multi-black hole
bound states around special loci in moduli space where
the central charges of the constituents become nearly aligned \cite{0206072,0702146}.
Thus, by considering a black hole whose dynamics in some 
region of the moduli space is described by a specific quiver quantum mechanics,
we can use our general Coulomb branch formula for multi-black hole bound states to 
parametrize the refined index in the corresponding quiver quantum 
mechanics  in terms of single-centered BPS invariants \cite{Manschot:2012rx}. 
For quivers without oriented loops, the single-centered BPS invariants are trivial,
and the Coulomb branch formula is completely explicit. 
We can then use this  
to establish the equivalence of the Coulomb branch formula for such quivers with explicit formulae for the cohomology of the moduli space
of stable quiver representations known in the mathematical literature \cite{MR1974891,Joyce:2004,ReinekeStoppaWeist,Moz:2012}.

Before explaining our results, let us briefly review the relation between quiver quantum mechanics
and multi-centered black holes  \cite{Manschot:2012rx}.
$\NN=4$ supersymmetric 
quiver quantum mechanics can be obtained by dimensionally
reducing an $\NN=1$ supersymmetric gauge theory in 3+1 dimensions\footnote{As an aside it should be noted that $\NN=4$ supersymmetric 
quiver quantum mechanics is also useful in computing the spectrum of
BPS states in $\NN=2$ supersymmetric gauge theories in 3+1 dimensions\cite{1102.1729,1107.0723}. 
Indeed, the BPS spectrum of many gauge 
theories can be understood in the language of quiver representations \cite{1112.3984}. 
For example in the context of $\NN=2$ supersymmetric pure SU(2) gauge theory,
the role of single-centered `black holes' is played by 
the monopole and the dyon in Seiberg-Witten theory,
which are stable for all values of the moduli and whose bound states generate the complete
BPS spectrum. } -- containing
vector multiplets in the adjoint representation of the gauge group $G=\prod_{i=1\dots K} U(N_i)$ 
and chiral multiplets in bi-fundamental representations of $U(N_i)\times U(N_j)$ --
 down to 0+1 dimensions.
The scalars coming from the chiral multiplets are called the
Higgs variables and those from the vector multiplets are called the Coulomb variables.
The vacuum moduli space of the Higgs variables at zero values of the
Coulomb variables  is equivalent to  
the moduli space $\cM$ of stable quiver representations (in short quiver moduli space),
where the stability condition is determined by the FI parameters $\zeta_1,\dots
\zeta_K$ for each $U(N_i)$ factor. BPS states are
in one-to-one correspondence with 
cohomology classes in $H^*(\cM;\IZ)$, and 
the angular momentum is identified with 
the component $J_3=(p-d)/2$ of the Lefschetz $SU(2)$ action on the total cohomology 
$H^*(\cM;\IZ)$. Thus 
the refined index is given by the `Poincar\'e-Laurent 
polynomial'\footnote{We use this terminology since $Q(\cM;y)$ 
differs from the usual Poincar\'e
polynomial $\sum_p b_p y^p$ by a $y\to -y$ transformation and a 
multiplicative factor of $(-y)^{-d}$.}
\be \label{epoly}
Q(\cM;y)\equiv \sum_{p} b_p(\cM)\, (-y)^{p-d},
\ee
 where $b_p(\cM)$ are the Betti numbers and $d$ is the complex dimension of $\cM$.
 But the same spectrum
may also be calculated by first integrating out all the Higgs variables and 
considering the effective theory for the Coulomb variables. The latter turns out to be given by the same quantum mechanical system as
that of multi-centered black holes in $\NN=2$ supersymmetric string theory,
upon identifying the charge vector $\gamma$ with the dimension vector 
$(N_1,\dots N_K)$ \cite{0206072}, and the DSZ product $\gamma_{ij}$ between the  
basis vectors $\gamma_i=(0,\dots,1,\dots,0)$, where the only non-vanishing
entry occurs in position $i$, with the number of arrows from the $i$-th to the
$j$-th node of the quiver. 
The Coulomb branch formula of \cite{1103.1887} can thus be used to express the 
Poincar\'e-Laurent polynomial  \refb{epoly} in terms of certain `single-centered BPS invariants' 
$\OmS(\alpha)$, which are (conjecturally) independent of the FI parameters and 
of the fugacity parameter $y$
\cite{Manschot:2012rx}. In simplest cases, $\OmS(\gamma)$ enters just as an 
additive constant in $Q(\cM;y)=\Omega(\gamma)$, and can be identified as the
Lefschetz singlet contribution to the total cohomology $H^*(\cM;\IZ)$ \cite{1205.5023,1205.6511,Manschot:2012rx,Lee2}. In general however, the 
single-centered BPS invariants $\OmS(\alpha)$  enter in  $Q(\cM;y)$  
in a more complicated fashion. 
It is a very interesting open problem to identify the corresponding
`absolutely stable' classes in $H^*(\cM;\IZ)$.

For quivers without oriented loops, the only non-vanishing single-centered BPS invariants are
those associated to the basis vectors $\gamma_i$, and for  
such vectors $\OmS(\gamma_i)$  
takes the value 1.
In that case, the Coulomb branch formula gives a completely explicit
result for $Q(\cM;y)$, which can be compared to known results in the mathematical literature.
In particular, for primitive dimension vector $\gamma$ (i.e. such that all $N_i$ are coprime), 
Reineke's formula \cite{MR1974891} gives another completely explicit result for $Q(\cM;y)$.
For Abelian quivers (i.e such that all $N_i$ are 
one), the Coulomb branch formula 
equates the Poincar\'e-Laurent polynomial $Q(\cM;y)$ with the Coulomb index 
$\gref(\{\gamma_1,\cdots \gamma_n\}; \{\zeta_1,\cdots \zeta_n\};y)$  
given in \eqref{eindexfinint}.
In \S\ref{sre1} we show the equivalence of the Reineke's formula 
and Coulomb branch 
results for Abelian quivers, 
generalizing previous arguments given in the context of wall-crossing \cite{1011.1258,1112.2515}.
For non-Abelian quivers with primitive dimension
vector, the equivalence of the Reineke's formula and Coulomb
branch results can be reduced to the Abelian case, by using the Abelianization property 
satisfied by Reineke's formula \cite{1112.2515,ReinekeStoppaWeist} (see
\eqref{MPSform} below).

For non-Abelian  
quivers with non-primitive vector, Reineke's formula no longer computes the Poincar\'e-Laurent polynomial 
$Q(\cM;y)$ of the quiver moduli space $\cM$ (which is singular due to marginal bound states),
but rather the 'stack invariant' $\gR$. It is conjectured in the mathematical literature that a bone-fide 
Poincar\'e-Laurent polynomial $Q(\cM;y)$ may be reconstructed from the stack invariants $\gR$ \cite{Joyce:2004, 0811.2435} (see Eq. \eqref{eq:inversestackinv} below), but it is unclear in general how to construct a smooth moduli space $\cM$ whose cohomology would agree 
with $\gR$. At any rate, using a
further property of Abelian stack invariants $\gRa$  
established in \cite{Moz:2012} (see \eqref{MRformula} below), we prove that the 
Coulomb branch formula agrees with the Poincar\'e-Laurent polynomial $Q(\cM;y)$ computed from 
Reineke's formula for the stack invariants. The result can be written as
\ben\label{erenonhiggsint}
Q\left(\cM;y\right) &=& \sum_{m|N_i \, \, \forall \, \, i} 
\frac{\mu(m)}{ m}  {y - y^{-1}\over y^m - y^{-m}}
\sum_{\{k^{(\ell)}_j\}\atop \sum_\ell \ell k^{(\ell)}_j=N_j/m}\!\!\!\!
\gref( \{ (\ell\gamma_j)^{k^{(\ell)}_j}\}; \{ (\ell \zeta_j)^{k^{(\ell)}_j}\}; y) \nonumber \\ &&
\,\prod_{j=1}^K 
\prod_\ell {1\over k^{(\ell)}_j!}\left( {y - y^{-1}\over \ell (y^\ell - y^{-\ell})}\right)^{k^{(\ell)}_j}
\ , 
\een
where $\mu(m)$ is the M\"obius function, $\{ (\ell\gamma_j)^{k^{(\ell)}_j}\}$ denotes
that we have $k^{(\ell)}_j$ nodes each carrying charge $\ell\gamma_j$ for $\ell\ge 1$
and $1\le j\le n$, and
it is understood that in computing $\gref$ whenever some $\alpha_{ij}$
vanishes we need to deform it away from that value to produce an Abelian
quiver without loop, and then use \refb{eindexfinint} for computing it.

For brevity we shall henceforth refer to Reineke's formula and
its generalizations as the Higgs branch formul\ae, since they compute directly the
cohomology of the Higgs branch moduli space. For quivers with loops, the Higgs branch
formula will refer to the result of any computation of the cohomology of the Higgs
branch moduli space, although no general formula is available in such cases. 

\subsection*{A {\sc Mathematica} package for quiver invariants}

It should be clear from the above that the Coulomb branch formula produces a parametrization of the Poincar\'e-Laurent polynomial of any quiver in terms of single-centered
BPS invariants in a completely combinatoric way. However, even for moderately complicated quivers the necessary computations quickly become tiresome, and are best implemented on
a computer. We have implemented the Coulomb branch formula as well as Reineke's formula 
for stack 
invariants (and many related other routines) in a  
mathematica package ``{\tt CoulombHiggs}'' available from arXiv and 
described in Appendix A, which we hope  will facilitate studies of 
single-centered BPS invariants. 
This package has been successfully tested
on the examples investigated in \cite{Manschot:2012rx} and many more.

\sectiono{A formula for the Coulomb index of multi-centered black holes
\label{sec_Coulomb}}

In this section, we establish a recursive algorithm for computing the Coulomb index 
$\gref(\{\alpha_1,\cdots $ $\alpha_n\};\{\dx_1,\cdots \dx_n\};y)$ for general charge
configurations $\alpha_i$, away from the walls of marginal stability in the space of FI 
parameters $\dx_i$. We start in \S\ref{sec_noloopnonzero}
with charge configurations 
for which there exists a possible ordering of the $\alpha_i$'s such that
\be \label{eord}
\alpha_{ij}\ge 0 \quad \hbox{for} \quad i\le j\, ,
\ee
and with all DSZ products $\alpha_{ij}$ non-zero.
It follows from the discussion in \S\ref{s1} that this corresponds to an Abelian
quiver without any oriented  loop.
For such systems we
obtain
the simple formula \eqref{efinal} for the indexed number $s(\sigma)$ that enters \refb{eindex}.  
In \S\ref{sec_noloopany}
we show that for the purpose of computing the total Coulomb index,
the same result can be used even when some of the $\alpha_{ij}$'s vanish.
In \S\ref{sec_genmulti}, we turn to general multi-centered black hole configurations 
for which the charges do not admit an ordering satisfying \refb{eord},
and establish an inductive formulae for computing $s(\sigma)$.
This is given by
$F(\{\ta_1,\cdots \ta_n\}, \{\cc_1,\cdots \cc_n\})$ in \refb{efinal1} with 
$\ta_i=\alpha_{\sigma(i)}$.
During this analysis we also derive a similar formula for the coefficient $s(\sigma)$ for
collinear scaling solutions\cite{Bena:2006kb,0702146} 
for which all the FI parameters vanish and the 
$\alpha_i$'s satisfy $\sum_{i<j}\alpha_{\sigma(i)\sigma(j)}=0$.  
The corresponding inductive formula for $s(\sigma)$, called 
$G(\alpha_{\sigma(1)},\cdots \alpha_{\sigma(n)})$, is given in
\refb{egrecur}. 
Combining these results and summing over all permutations yields a general algorithm 
for $\gref(\{\alpha_1,\cdots \alpha_n\};\{\dx_1,\cdots \dx_n\};y)$ with non vanishing
DSZ products, as summarized in \refb{eqforg}.

\subsection{Abelian quivers with no oriented loops and all $\alpha_{ij}\ne 0$}  \label{sec_noloopnonzero}

We start by considering the case of charge configurations which admit an ordering 
such that $\alpha_{ij}>0$ for $i<j$.  
A special case arises if all $\alpha_i$ are positive linear combinations of 
two charge vectors $\gamma_1,\gamma_2$ with $\langle \gamma_1 , \gamma_2 \rangle >0$,
as is the case for multi-centered configurations relevant for wall-crossing \cite{1011.1258}. 

Defining
\be
y_i = x_{i+1}-x_i, \qquad \tc_i = -\sum_{j=i+1}^n \cc_i
=\sum_{j=1}^i \cc_i, \quad \hbox{for $1\le i\le n-1$}\ ,
\ee
we can express \refb{ecou1} as
\be \label{ecou2}
W(\{y_i\})=-\sum_{i<j} \ta_{ij}\,  \ln \left(\sum_{k=i}^{j-1}  y_k \right)
+\sum_{i=1}^{n-1} \tc_i\, y_i \, .
\ee
This gives
\be \label{esign3}
{\p W\over \p y_k} = -\sum_{j=1}^{k} \sum_{\ell= k+1}^n \, \ta_{j\ell}\, 
{1\over \sum_{i=j}^{\ell-1} y_i}
+\tc_k \quad \hbox{for $1\le k\le n-1$}\, .
\ee
Each $y_i$ takes value from 0 to $\infty$, but by an appropriate coordinate
transformation $u_i=f(y_i)$ we can bring the range to $0\le u_i<1$ for each $i$.
We can regard the space spanned by the $u_i$'s a unit box.
Our goal is to examine the condition under which $W$ has an extremum with respect
to the $y_i$'s in the interior of this box, \i.e.\ there is a solution to the equation
$\p W/\p y_k=0$ for every $k$. 

To this aim,
let us now consider the following deformation of the $\ta_{ij}$'s:
\be \label{edeform1}
\ta_{i.i+1}\to \ta_{i,i+1} \, \, \forall \, i, \qquad 
\ta_{ij}\to \lambda \, \ta_{ij} \, \, \, \, \hbox{for $|i-j|\ge 2$} \, , \qquad 0\le\lambda\le 1\, .
\ee
In the  limit $\lambda\to 0$, only $\ta_{i,i+1}$'s remain non-zero and \refb{esign3}
takes the simple form
\be \label{edeform2}
{\p W\over \p y_k} = - {\ta_{k,k+1}\over y_k}+ \tc_k\, .
\ee
Thus the set of equations
$\p W/\p y_k=0$ has a solution in the range $0<y_k<\infty$ if and only if
\be \label{esign}
{\rm sign}(\ta_{k,k+1}) ={\rm sign} (\tc_k) \quad \hbox{for $1\le k\le n-1$}\, .
\ee
Furthermore the sign of the
Hessian of $W$ at this solution is easily determined to be
\be \label{esign2}
\prod_{k=1}^{n-1} \, {\rm sign}(\ta_{k,k+1}) \, .
\ee
These results can be summarized by saying that for quivers without
loops the coefficient $s(\sigma)$ of
$y^{\sum_{i<j}\ta_{ij}}$ associated with 
a given permutation $\{\ta_1,\cdots \ta_n\}$ is
given, for $\lambda=0$, by
\be \label{efinal}
F_0(\{\ta_1,\cdots \ta_n\}, \{\cc_1,\cdots \cc_n\})
= \prod_{k=1}^{n-1} \Theta(\ta_{k, k+1} \, \tc_k) (-1)^{\sum_{k=1}^{n-1}
\Theta(-\ta_{k,k+1})}\, ,
\ee
where $\Theta(x)$ is the Heaviside function.

Now consider changing $\lambda$ from 0 to 1. During this deformation
new extrema can appear and disappear in pairs in the interior of the box, but
since they are weighted by the sign of the Hessian, the result is
unaffected. The other possibility is that new extrema can emerge 
from (or disappear into) the boundary of the box.
This happens when a subset of the $y_i$'s approach  0 and/or infinity.
Since $\p W/\p y_k$ approaches the constant
$\tc_k$ as $y_k\to\infty$  irrespective of the values of the other $y_i$'s,
it is clear that away from the marginal stability
walls $\tc_k=0$, none of the $y_i$'s
can approach infinity. Thus the only
possible boundary component where extrema of $W$ can appear or disappear
is where a subset of the $y_i$'s vanish.

Let us suppose that as  $\lambda$ is increased from 0 to 1, such a phenomenon takes
place at some value $\lambda=\lambda_c$. If a new extremum appears at $\lambda_c$
then for $\lambda$ slightly above $\lambda_c$, there will be an extremum of $W$
where a subset of the $y_i$'s are small, corresponding to a subset of the centers being
close to each other. If on the other hand an extremum disappears as
$\lambda$ approaches $\lambda_c$ from below, then such a configuration exists
for $\lambda$ slightly below $\lambda_c$. These correspond to onset or disappearance
of scaling solutions\cite{Bena:2006kb,0702146}, with $\lambda=\lambda_c$ 
being the point at which the scaling
solution becomes collinear.

Now if the quiver corresponding to the
original $\alpha_{ij}$'s had no oriented  loop  then 
neither does the quiver associated with the deformed $\alpha_{ij}$'s,
since the signs of all the $\alpha_{ij}$'s are 
preserved under
the deformation \eqref{edeform1}. This implies that the deformed quiver cannot have a scaling
configuration, and hence, as we deform $\lambda$ from 0 to 1, no extremum of $W$
can emerge from or disappear into the boundary. Thus \refb{efinal} gives
the correct contribution to $s(\sigma)$ even at $\lambda=1$.
Using \refb{eindex} we now get
\be \label{eindexfin}
\begin{split}
 \gref(\{\alpha_1,\cdots \alpha_n\}; \{\dx_1,\cdots \dx_n\};y) 
=& (-1)^{n-1+\sum_{i<j} \alpha_{ij}} (y-y^{-1})^{-n+1} \\
 \sum_\sigma 
\prod_{k=1}^{n-1}  \Theta\left(\alpha_{\sigma(k), \sigma(k+1)} \, \sum_{i=1}^k c_{\sigma(k)} \right) &
(-1)^{\sum_{k=1}^{n-1}
\Theta(-\alpha_{\sigma(k),\sigma(k+1)})}\, 
y^{\sum_{i<j} 
\alpha_{\sigma(i)\sigma(j)}} \, . 
\end{split}
\ee

\subsection{Abelian quivers with no oriented loops but some $\alpha_{ij}=0$} \label{sec_noloopany}

We now turn to the case of quiver without oriented loops, but for which some of the
$\alpha_{ij}$'s vanish. For this we first deform all the  vanishing $\alpha_{ij}$'s to non-zero
values in such a way  that the deformed quiver does not have  any oriented loop. To see that this is
always possible, let us carry out the deformation one link at a time.
We begin with the original quiver without oriented loop and make one of
the vanishing $\alpha_{ij}$'s non-zero. If this leads to a quiver with
an oriented loop, then there exists some component $C$ of the original quiver, which,
together with the new link, gives rise to a quiver with oriented loop. If so let us flip
the sign of $\alpha_{ij}$ of the deformed link. In this case $C$ together with the added link
no longer forms an oriented loop. Suppose there were another component $C'$ of the original
quiver, which combined with the new link would now form an oriented loop. 
Then $C+C'$  
would form an oriented loop in the original quiver, whih contradicts our assumption.
Thus, by choosing the sign of $\alpha_{ij}$ of the deformed link we can ensure
that the new quiver also does not have any oriented loop. We can now repeat the argument
and show that all the vanishing $\alpha_{ij}$'s can be made non-zero and for appropriate
choice of sign of the deformed $\alpha_{ij}$'s the new quiver does not have any oriented
 loop. Thus we can compute its index by our earlier formula \refb{eindexfin}. 

We shall now argue that the index of the original quiver can be obtained by taking the
limit of the index of the deformed quiver in which the deformation parameters go to zero.
For this we shall work with the total index \refb{eindex} rather than a given permutation.
We shall use the original charges $\alpha_i$ and
use their locations $z_i$ -- related to the $x_i$ by $x_i=z_{\sigma(i)}$ -- as the independent
variables. The $z_i$'s satisfy \refb{eor1}.

Let us now consider the effect of taking $\alpha_{pq}$ to 0 for some
specific $p$, $q$. For any extremum of $W$ at which the locations of $\alpha_p$ and 
$\alpha_q$ remain at finite separation, this limit has no drastic effect and the
contribution to the index from such extrema at $\alpha_{pq}=0$ is the same as what we
get by taking the $\alpha_{pq}\to 0$ limit. Thus we only have to examine the fate of
the critical points for which the locations of $\alpha_p$ and $\alpha_q$ approach each other
in the $\alpha_{pq}\to 0$ limit, as generically such critical points will disappear for
$\alpha_{pq}=0$. For such solutions we can replace $z_q$ by $z_p$ in
\refb{eor1} except in the $\alpha_{pq}/|z_p-z_q|$ terms, and
express \refb{eor1} as  
\be \label{eor2}
\begin{split}
-\sum_{i=1\atop i\ne k, p, q}^n {\alpha_{ik}\over |z_i - z_k|} 
- {\alpha_{pk}+\alpha_{qk}\over |z_p-z_k|} 
- \dx_k =& 0\, \quad \hbox{for $1\le k\le n$, $k\ne p,q$} \\
-\sum_{i=1\atop i\ne  p, q}^n {\alpha_{ip}\over |z_i - z_p|}
- {\alpha_{qp}\over |z_q - z_p|} - \dx_p=&0 \ ,\qquad
-\sum_{i=1\atop i\ne  p, q}^n {\alpha_{iq}\over |z_i - z_p|}
- {\alpha_{pq}\over |z_q - z_p|} - \dx_q=0\, .
\end{split}
\ee
By adding and subtracting the last two 
equations we get 
\be \label{eor3}
\begin{split}
-\sum_{i=1\atop i\ne k, p, q}^n {\alpha_{ik}\over |z_i - z_k|} 
- {\alpha_{pk}+\alpha_{qk}\over |z_p-z_k|} 
- \dx_k = & 0\,  \quad \hbox{for $1\le k\le n$, $k\ne p,q$}\\
-\sum_{i=1\atop i\ne  p, q}^n {\alpha_{ip}+\alpha_{iq}\over |z_i - z_p|}
- \dx_p - \dx_q=&0 \ ,\qquad 
-\sum_{i=1\atop i\ne  p, q}^n {\alpha_{ip}-\alpha_{iq}\over |z_i - z_p|}
+2  {\alpha_{pq}\over |z_q - z_p|} - \dx_p+ \dx_q=0\, .
\end{split}
\ee
The first set of equations and the second equation together correspond to
the equilibrium configuration of $n-1$ charges in which the charges
$\alpha_p$ and $\alpha_q$ have merged to form a charge $\alpha_{p}+\alpha_q$ 
and
the corresponding FI parameters have been added. The last
equation can be interpreted as an equation for $z_q-z_p$. The existence of a 
solution to this equation requires 
\be \label{econd}
{\rm sign} \left(\dx_q- \dx_p -\sum_{i=1\atop i\ne  p, q}^n {\alpha_{ip}-\alpha_{iq}\over |z_i - z_p|}
\right) = {\rm sign} \, \alpha_{qp}\, .
\ee
When this condition is satisfied then
$z_q-z_p$ is of order  $\alpha_{pq}$ for small $\alpha_{pq}$. 
On the other hand 
when $\alpha_{pq}=0$, the last equation in \refb{eor3} generically 
has no solution since the left hand side of the equation becomes independent
of $z_q$ and all the $z_i$'s for $i\ne q$ 
are already fixed by the other equations.
This shows that the critical points of $W$ associated with solutions
to \refb{eor3} disappear at $\alpha_{pq}=0$. This could give rise
to a discontinuity in the index at $\alpha_{pq}=0$.
The important point to note
however is that even if \refb{econd}  is satisfied, the solutions
to the last equation in \refb{eor3} always occur in pairs, related by a reversal of
the sign of $z_p-z_q$. The exponent of $y$ in \refb{eindex1} remains unchanged
under this exchange in the $\alpha_{pq}\to 0$ limit since  this exchange
only flips the sign of the coefficient of $\alpha_{pq}$ in the exponent. 
Finally it is easy to check that $s(\sigma)$ changes sign under this exchange. Thus
the contribution from this pair of solutions cancel and we get a smooth $\alpha_{pq}\to 0$
limit. Repeating this analysis for the other $\alpha_{pq}$'s we see that the index associated
with the original quiver can be obtained as the limit of the index associated with the deformed
quiver. 

Note that the above argument breaks down if the solution to the
equations in the first two lines of \refb{eor3} automatically 
satisfy 
\be  
\dx_q- \dx_p -\sum_{i=1\atop i\ne  p, q}^n {\alpha_{ip}-\alpha_{iq}\over |z_i - z_p|}=0\, .
\ee
This happens for example when $\alpha_p$ and $\alpha_q$ are  parallel so that 
$\alpha_{ip}/\alpha_{iq}
=|\alpha_p|/|\alpha_q|=\dx_p/\dx_q$. In this case at $\alpha_{pq}=0$
there is a solution to \refb{eor1} at $z_p=z_q$, obtained by solving the equations in
the first two lines of \refb{eor3}. Now consider the case when $\alpha_{pq}$
is deformed away from 0. In this case in order to look for a solution to the last equation in
\refb{eor3} where $z_p$ and $z_q$
are close to each other we can no longer set $z_p=z_q$ in the regular terms from the beginning,
but must keep terms of order $(z_p-z_q)$ in the last equation. If we call this term
$A(z_p-z_q)$ for some constant $A$ then we can express this equation as
\be \label{exp}
A(z_p-z_q) + 2  {\alpha_{pq}\over |z_q - z_p|} = 0\, .
\ee
This equation is no longer invariant under a change of sign of $z_p-z_q$, and in fact has
a solution only for one sign of $z_p-z_q$ irrespective of the sign of $\alpha_{pq}$.  
In the $\alpha_{pq}\to 0$ limit
this solution smoothly continues to
the solution with $z_p=z_q$ at $\alpha_{pq}=0$. Thus we again see that the 
$\alpha_{pq}\to 0$
limit is smooth, and agrees with the result for $\alpha_{pq}=0$.

\subsection{Generic Abelian quivers with all $\alpha_{ij}$
non-zero \label{sec_genmulti}}

We shall now consider 
a generic multi-centered black hole configuration with all $\alpha_{ij}$ non-zero, 
but  whose associated quiver may possess  oriented loops. In this case
we need to take into account possible contributions from scaling solutions. 
Our goal
in this section will be to compute $\gref$ for such configurations.

\subsubsection{An inductive formula for the index of collinear solutions\label{sec_CoulombF}}

We proceed as in \S\ref{sec_noloopnonzero} and consider the deformation 
\refb{edeform1}. It is clear that at $\lambda=0$ the contribution from a given permutation
$\sigma$ will be given by \refb{efinal}. Thus we need to investigate the
total change in this contribution as $\lambda$ changes from 0 to 1.
As discussed in \S\ref{sec_noloopnonzero} these changes could come from values of $\lambda$ at which
a set $A$ of neighbouring centers come close to each other. This can happen if the total
angular momentum carried by this set of centers vanish,
\be \label{ecalpha}
\sum_{i,j\in A; i<j } \ta'_{ij} =0\ ,
\ee 
where $\ta'_{ij}$ denotes the deformed $\ta_{ij}$. To see this, note that in the
limit where all $y_i$ for $i,i+1\in A$ approach zero,  
the part of the superpotential \eqref{ecou2} involving the $y_i$'s 
becomes quasi-homogeneous,
\be
W( \{\lambda y_i\}) \sim  W( \{y_i\}) - \sum_{i,j\in A; i<j } \ta'_{ij} \, \log \lambda\ .
\ee
Differentiating with respect to $\lambda$ and using $\partial W/\partial y_i=0$ implies \eqref{ecalpha}.
Since the set $A$ must contain at least three elements there are at most 
$(n-2)(n-1)/2$ possible sets $A$, given by the $(n-2)(n-1)/2$ possible ways 
of choosing the beginning and the end of the set. Correspondingly there are  at most
$(n-2)(n-1)/2$ possible 
values $\lambda_A$ of the deformation parameter $\lambda$
where such collinear scaling configurations
can arise.
Using \refb{edeform1}, the condition \refb{ecalpha}  becomes 
a linear equation in $\lambda_A$,
\be \label{erange}
\lambda_A \, \sum_{i,j\in A; i\le j-2} \ta_{ij}+ \sum_{i\in A, i+1\in A} \ta_{i,i+1} = 0\, .
\ee
The index can jump across $\lambda=\lambda_A$ if $\lambda_A$ 
lies between 0 and 1. 
This is so if and only if the left hand side of \refb{erange}
has opposite signs at $\lambda_A=0$ and 1, i.e. 
\be \label{eneweq}
\left(\sum_{i\in A, i+1\in A} \ta_{i,i+1}\right) \left( \sum_{i,j\in A, i<j} \ta_{ij}\right) < 0\, .
\ee
If $F(\{\ta_1,\cdots \ta_n\}, \{\cc_1,\cdots \cc_n\})$ denotes the coefficient $s(\sigma)$
of $y^{\sum_{i<j}\ta_{ij}}$ associated with a given permutation, then we have
\be \label{efinal1}
F(\{\ta_1,\cdots \ta_n\}, \{\cc_1,\cdots \cc_n\})=
F_0(\{\ta_1,\cdots \ta_n\}, \{\cc_1,\cdots \cc_n\})
+ \sum_A \Delta F_A \, ,
\ee
where $F_0$ is given by \refb{efinal}, and $\Delta F_A$ is the jump across
the  critical point $\lambda_A$.
Our goal will be to compute the expression for $\Delta F_A$.

Let us suppose that set $A$ consists of the integers
$k,k+1,\cdots\ell$. 
We shall examine the configuration close to the critical point 
by taking 
\be \label{edefeps}
\sum_{r,s\atop k\le r<s\le \ell} \ta'_{rs} = \epsilon
\ee
for some small number $\epsilon$. 
We now define $z_s$ via 
\be
\label{defxyz}
x_s = x_k + y\, z_s \quad \hbox{for $k\le s \le \ell$}\, ,\quad z_k\equiv 0\ , \quad 
z_\ell\equiv 1\ ,
\ee
and use $x_1, \cdots x_k, x_{\ell+1}, \cdots x_n, y, z_{k+1},\cdots z_{\ell-1}$
as independent variables. Then the relevant equations are given by  extremizing
\ben \label{etot}  
W &=& - \sum_{i,j\atop i<j; i,j< k \, or\, \ge \ell+1} \ta'_{ij} \ln (x_j - x_i)
- \sum_{i=1}^{k-1} \sum_{s=k}^{\ell} \ta'_{i s} \ln (x_k + y z_s  - x_i)
\nonumber \\ 
&& - \sum_{i= \ell+1}^n  \sum_{s=k}^{\ell} \ta'_{si} \ln(x_i - x_k - y z_s)
-\sum_{s,r\atop k\le s<r\le \ell} \ta'_{sr} \ln (y (z_r-z_s))
 \nonumber \\ &&
-   \sum_{i\atop i< k \, or\, i\ge \ell+1}\,  \cc_i x_i - \left(\sum_{s=k}^\ell \cc_s\right) x_k
- y \sum_{s=k+1}^{\ell} \cc_s z_s \, . 
\een
We shall be examining an extremum of
$W$ for which $y$ is small, of order $\epsilon$. In this case
the extrema of $W$ with respect to $x_1,\cdots x_k$, $x_{\ell+1},\cdots x_n$
can be obtained by extremizing
\ben \label{eone}
W_1 &=& - \sum_{i,j\atop  i<j; i,j < k \, or\,\ge \ell+1} \ta'_{ij} \ln (x_j - x_i)
- \sum_{i=1}^{k-1} \left(\sum_{s=k}^{\ell} \ta'_{i s}\right) \ln (x_k   - x_i)
  \nonumber \\ &&
- \sum_{i= \ell+1}^n  \left(\sum_{s=k}^{\ell} \ta'_{si}\right) \ln (x_i - x_k)
  -  \sum_{i\atop i< k \, or\, i\ge \ell+1}\,   \cc_i x_i 
  - \left(\sum_{s=k}^\ell  \cc_s\right) x_k . 
\een
The existence of an extremum of $W_1$
is equivalent to the existence of 
a collinear configuration with $n-\ell+k$ centers with charges
$\ta'_1,\cdots \ta'_{k-1}, \sum_{s=k}^\ell \ta'_s, \ta'_{\ell+1}, \cdots \ta'_n$
and FI parameters $ \cc_1,\cdots  \cc_{k-1}, \sum_{s=k}^\ell  \cc_s,
 \cc_{\ell+1},\cdots  \cc_n$, situated at $x_1,\cdots x_{k-1}, x_k, x_{\ell+1},\cdots x_n$.
Extremization of \refb{etot} with respect to the parameters $z_s$ for $k+1\le s\le \ell-1$
can be obtained by extremizing
\be \label{etwo}  
W_2 = 
-\sum_{s,r\atop k\le s<r\le \ell} \ta'_{sr} \ln (z_r-z_s) \, .
\ee
The existence of an extremum of $W_2$ is equivalent to
the existence of  a collinear scaling configuration of $\ell-k+1$ centers with charges
$\ta'_k,\cdots \ta'_\ell$ and zero FI parameters, with the locations
of the centers being at $z_k=0$, $z_{k+1}$, $\cdots$, $z_{\ell-1}$ and 
$z_\ell=1$. Finally $y$ can be
obtained by extremizing\footnote{Since $y$ is small, we have expanded the terms in $W$ 
which are non-singular in the $y\to 0$ limit to first order in $y$.}
\be \label{ethree}
W_3 = - \sum_{s,r\atop k\le s<r\le \ell} \ta'_{sr} \ln y
- y\sum_{i=1}^{k-1} \sum_{s=k+1}^{\ell} \ta'_{i s} {z_s\over x_k-x_i}
+ y\sum_{i=\ell+1}^n  \sum_{s=k+1}^{\ell} \ta'_{si} {z_s\over x_i-x_k} 
 - y\sum_{s=k+1}^\ell  \cc_s z_s\, .
\ee
Using \refb{edefeps} this gives  
\be
-{\epsilon \over y} - \sum_{i=1}^{k-1} \sum_{s=k+1}^{\ell} \ta'_{i s} {z_s\over x_k-x_i}
+ \sum_{i=\ell+1}^n  \sum_{s=k+1}^{\ell} \ta'_{si} {z_s\over x_i-x_k} 
 - \sum_{s=k+1}^\ell  \cc_s z_s = 0\, .
 \ee
 For generic $\ta_{ij}$,
a solution to this equation with positive $y$ exists only for one choice of sign
 of $\epsilon$. We shall assume that we have taken the sign of $\epsilon$ to be such
 that the solution exists. 
Let $\eta_A$ denote
a quantity which takes value $1$ ($-1$) if the solution exists for $\lambda$ above
(below) the critical value $\lambda_A$ given in \eqref{erange}.

Let us now compute the Hessian at this critical point. From \refb{eone}-\refb{ethree}
it is clear that the second derivative of $W$ with respect to all the variables except $y$
remain finite, and we have 
$\p^2 W /\p y^2\sim \sum_{s,r\atop
k\le s<r\le \ell} \ta'_{sr}/y^2\sim \epsilon/y^2$.  
Since this is large for $y\sim\epsilon$ and all other
second derivatives of $W$ remain finite, the full determinant will be given by the product
of $\p^2 W /\p y^2$ and the determinant of the Hessian involving the rest of the variables.
Furthermore we see from \refb{etot} that $\p^2 W/ \p x^i \p z_s$ goes to zero as $y\to 0$.
Thus the Hessian of $W$ with respect to $x_i$'s and $z_s$'s factorizes into 
the product of the
Hessian of $W_1$ with respect to
$x_1,\cdots x_k, x_{\ell+1},\cdots x_n$ and the Hessian of $W_2$ with respect to
$z_{k+1},\cdots z_{\ell-1}$.
In our notation the sign of the Hessian of $W_1$ with
respect to $x_1,\cdots x_k, x_{\ell+1},\cdots x_n$ is given by
$F(\{\ta'_1,\cdots \ta'_{k-1}, \ta'_k+\cdots \ta'_\ell, \ta'_{\ell+1},\cdots \ta'_n\},
\{\cc_1,\cdots \cc_{k-1}, \cc_k+\cdots \cc_\ell, \cc_{\ell+1},\cdots \cc_n\})$.
Let   
\be
G(\ta'_k, \cdots \ta'_\ell) = {\rm sign}\left(  \det_{k+1\leq i,j\leq \ell-1} (\p_{z_i}\p_{z_j} W_2) \right)
\ee
be the sign of the Hessian of $W_2$ with respect to
$z_{k+1},\cdots z_{\ell-1}$ when the corresponding scaling solution exists; otherwise
we take $G(\ta'_k, \cdots \ta'_\ell)=0$.\footnote{As usual if there are more than
one solutions then we add their contributions.}  
Then we can write
\ben \label{edeltafs}
\Delta F_A &=& F(\{\ta'_1,\cdots \ta'_{k-1}, \ta'_k+\cdots \ta'_\ell, \ta'_{\ell+1},
\cdots \ta'_n\},
\{\cc_1,\cdots \cc_{k-1}, \cc_k+\cdots \cc_\ell, \cc_{\ell+1},\cdots \cc_n\})\nonumber \\ &&
\times \, \, \eta_A \, {\rm sign}(\epsilon) \, G(\ta'_k, \cdots \ta'_\ell)\,
\Theta\Big(-
\big(\sum_{i=k}^{\ell-1} \ta_{i,i+1}\big) \big( \sum_{i,j\atop k\le i<j \le \ell} \ta_{ij}\big) 
\Big)\, ,
\een
where the last factor imposes the constraint \refb{eneweq}.
Now it  
follows from \refb{edeform1} that at $\lambda=\lambda_A + \delta\lambda$,
\be \label{ekey}
\epsilon = \sum_{r,s\atop k\le r<s\le \ell} \ta'_{rs} =
\delta\lambda \, \sum_{i,j=k\atop
i\le j-2}^\ell \ta_{ij} \, .
\ee
Suppose the solution exists for $\delta\lambda>0$. Then we have $\eta_A=1$ and
we see from \refb{ekey} that sign($\epsilon$)=sign($\sum_{i,j=k\atop
i\le j-2}^\ell \ta_{ij}$). On the other hand if the solution exists for $\delta\lambda<0$ then
we have 
$\eta_A=-1$ and
 sign($\epsilon$)=$-$sign($\sum_{i,j=k\atop
i\le j-2}^\ell \ta_{ij}$). Thus in either case
\be
\eta_A \, {\rm sign}(\epsilon) = {\rm sign}\Big(\sum_{i,j=k\atop
i\le j-2}^\ell \ta_{ij}\Big) \, .
\ee
Substituting this into \refb{edeltafs} we get
\ben \label{edeltafs1}
\Delta F_A &=& F(\{\ta'_1,\cdots \ta'_{k-1}, \ta'_k+\cdots \ta'_\ell, \ta'_{\ell+1},
\cdots \ta'_n\},
\{\cc_1,\cdots \cc_{k-1}, \cc_k+\cdots \cc_\ell, \cc_{\ell+1},\cdots \cc_n\})\, \nonumber \\ &&
\times \, \, G(\ta'_k, \cdots \ta'_\ell) \,  \, {\rm sign}\left(\sum_{i,j=k\atop
i\le j-2}^\ell \ta_{ij}\right) \, 
\Theta\left(-
\left(\sum_{i=k}^{\ell-1} \ta_{i,i+1}\right) \left( \sum_{i,j\atop k\le i<j \le \ell} \ta_{ij}\right) 
\right)\, . \nonumber \\
\een
Note that a special case of \refb{edeltafs1} is $k=1$, $\ell=n$ in which case
the $F$ on the right hand side of this equation is 
$F(\ta'_1+\cdots \ta'_n; \cc_1+\cdots \cc_n)=1$.
If we can compute $G(\ta'_k, \cdots \ta'_\ell)$, then we
can use \refb{efinal1} and \refb{edeltafs1} to compute the function $F$ recursively.
Once we know how to compute $F$, the Coulomb index can be computed as
\be \label{eqforg} 
\begin{split}
 \gref&(\{\alpha_1,\cdots \alpha_n\}; \{\dx_1,\cdots \dx_n\};
y) \\
=& (-1)^{n-1+\sum_{i<j}\alpha_{ij}}
(y-y^{-1})^{-n+1}  \sum_{\sigma}
F\left(\{\alpha_{\sigma(1)},\cdots \alpha_{\sigma(n)}\}; \{\dx_{\sigma(1)},\cdots
\dx_{\sigma(n)}\}
\right) y^{\sum_{i<j} \alpha_{\sigma(i)\sigma(j)}}\, , 
\end{split}
\ee
where the sum runs over all permutations $\sigma$.

\subsubsection{An inductive formula for the index of scaling collinear solutions
\label{sec_CoulombG}}

We now turn to the computation of $G(\ha_1,\cdots \ha_m)$, the indexed number
of critical points of the superpotential
\be 
\label{ewhW}
\wh W = -\sum_{i,j\atop 1\le i<j\le m} \, \ha_{ij} \, \ln (z_j - z_i)\ , \quad 
\sum_{i,j\atop 1\le i<j\le m} \, \ha_{ij} =0\, ,
\ee
in the range $z_j>z_i$ for $j>i$. This coincides with \eqref{etwo} under the identifications $\{\ha_1,\cdots \ha_m\}=
\{\ta'_k, \cdots \ta'_\ell\}$, and obvious redefinitions of $z_i$. The invariance 
of the superpotential
\eqref{ewhW} under both translation and rescaling of the $z_i$'s,
must be `gauge fixed' before counting critical points. These invariances
were fixed by the conditions $z_1=0, z_m=1$ in \eqref{defxyz};
however in order to compute $G(\ha_1,\cdots \ha_m)$ inductively,
it will be more convenient to choose a different gauge $z_1=0$, $z_{m-1}=1$.

Let us now consider the deformation 
\be \label{edefa}
\ha_{im}\to \mu \, \ha_{im} \quad \hbox{for $i=1,2,\cdots m-1$}, 
\quad \ha_{m-3,m-1} \to \ha_{m-3,m-1}
+ (1-\mu) \sum_{i=1}^{m-1} \ha_{im}\, ,
\ee
so that the deformed $\ha_{ij}$'s (called $\ha'_{ij}$) continue to satisfy
$\sum_{i<j}\ha'_{ij}=0$. In the limit $\mu\to 0$, we can treat the $m$-th center as
a probe in the background of other centers. From the behaviour of $\wh W$ as a function
of $z_m$ in the two limits, $z_m\to z_{m-1}$ and $z_m\to\infty$, we conclude that the
solution exists if and only if the $m-1$ centered scaling solution with 
$\ha_{ij}$ for $1\le i,j\le m-1$ given by \refb{edefa} for $\mu=0$ exists, and
furthermore
\be \label{edefb}
{\rm sign} (\ha_{m-1,m}) = - {\rm sign} \left(\sum_{i=1}^{m-1}\ha_{im}\right)\, .
\ee
Finally the sign of the Hessian associated with the configuration in the $\mu\to 0$ 
limit, after adding up the contribution from all critical points in the range  
$z_j>z_i$ for $j>i$, is
\be \label{edefc}
{\rm sign} (\ha_{m-1,m})\, G(\ca_1, \cdots \ca_{m-1})  \, ,
\ee
where $\ca_i$ for $1\le i\le m-1$ denote the deformed charges at $\mu=0$.
Using \refb{edefa}-\refb{edefc} we get 
\be \label{egrecur}
G(\ha_1,\cdots \ha_m) = (-1)^{1+\Theta(\ha_{m-1,m})} \Theta\left(-
\ha_{m-1,m} \sum_{i=1}^{m-1}\ha_{im}
\right) \, G(\ca_1, \cdots \ca_{m-1}) + \sum_B \Delta G_B\, ,
\ee
where $\Delta G_B$ denotes the jump in $G$ during the deformation from $\mu=0$
to $\mu=1$ across the various critical points $\mu_B$ where a subset $B$
of the charges
can form scaling solutions.\footnote{Due to scale invariance 
a configuration where a subset of the centers get infinitely separated from
the others is equivalent to a configuration where a subset of the centers come 
together.}
Since the deformations involve the $\ha_{im}$ and
$\ha_{m-3,m-1}$, new scaling solutions must involve either the $m$th center, or
both the $(m-1)$'th and $(m-3)$'th center.
Three kinds of scaling solutions can be encountered  
during the deformation: 
\begin{enumerate}
\item The scaling configuration involves the charges $\ha_{m-2}$, $\ha_{m-1}$ and
$\ha_m$. In this case we need
\be \label{enewforce}
\ha'_{m-2,m-1} + \ha'_{m-1,m} + \ha'_{m-2,m}=0\, ,
\ee
which requires
\be \label{eama}
\ha_{m-2,m-1} + \mu_B (\ha_{m-1,m} + \ha_{m-2,m})=0\, .
\ee
\item The scaling configuration  involves charges $\ha_k,\cdots \ha_m$ for $2\le k\le (m-3)$. In this case
we require
\be \label{especial}
\sum_{i,j\atop k\le i<j\le m} \ha'_{ij}=0 
\ee
which translates to
\be \label{eamb}
\mu_B\sum_{i=k}^{m-1} \ha_{im} + \sum_{i,j\atop
k\le i<j\le m-1} \ha_{ij}
+ (1-\mu_B) \sum_{i=1}^{m-1} \ha_{im} =0\, .
\ee
\item The scaling configuration 
involves charges $\ha_k,\cdots \ha_{m-1}$ for $2\le k\le (m-3)$. In this case
we require
\be 
\sum_{i,j\atop k\le i<j\le m-1} \ha'_{ij}=0 
\ee
which translates to
\be \label{eamc}
 \sum_{i,j\atop k\le i<j\le m-1} \ha_{ij}
+ (1-\mu_B) \sum_{i=1}^{m-1} \ha_{im} =0\, .
\ee
\end{enumerate}
For each of these cases the computation of $\Delta G_B$ follows the procedure used
for computing $\Delta F_A$ earlier. We shall quote the final results generalizing
\refb{edeltafs1}:
\begin{enumerate}
\item For $\mu$ satisfying \refb{eama} for $m>4$ we have
\ben
\label{escal1}
\Delta G_B &=& G(\ha_1', \cdots \ha'_{m-3}, \ha'_{m-2}+\ha'_{m-1}+\ha'_{m}) \times
G(\ha'_{m-2}, \ha'_{m-1}, \ha'_m) \,   \nonumber \\
&& \times \, {\rm sign} \left(\ha_{m-1,m} + \ha_{m-2,m}
\right) \, \Theta\left(- 
\left(\ha_{m-2,m-1} +\ha_{m-1,m} + \ha_{m-2,m}\right)
\ha_{m-2,m-1}
\right)\, . \nonumber \\
\een
The case $m=4$ requires special attention and will be discussed later.
\item For $\mu$ satisfying \refb{eamb} with $k>2$ we have
\ben
\label{escal2}
\Delta G_B &=& G\left(\ha_1', \cdots \ha'_{k-1}, \sum_{i=k}^m \ha'_{i}\right) \times
G(\ha'_{k}, \ha'_{k+1}, \cdots, \ha'_m) \, {\rm sign} \left(-\sum_{i=1}^{k-1} \ha_{im}
\right)\nonumber \\
&& \times \Theta\left(-
\left(\sum_{i,j\atop k\le i<j\le m} \ha_{ij} \right)
\left( \sum_{i,j\atop k\le i<j\le m-1} \ha_{ij}
+  \sum_{i=1}^{m-1} \ha_{im} \right)
\right)\, .
\een
The case $k=2$ requires special treatment and will be discussed
below.
\item For $\mu$ satisfying \refb{eamc} we have
\ben
\label{escal3}
\Delta G_B &=& G\left(\ha_1', \cdots \ha'_{k-1}, \sum_{i=k}^{m-1} \ha'_{i}, \ha'_m\right) \times
G(\ha'_{k}, \ha'_{k+1}, \cdots, \ha'_{m-1}) \, {\rm sign} \left(-\sum_{i=1}^{m-1} \ha_{im}
\right) \nonumber \\ &&
\times \Theta\left(-
\left( \sum_{i,j\atop k\le i<j\le m-1} \ha_{ij}
+  \sum_{i=1}^{m-1} \ha_{im}\right)
\left( \sum_{i,j\atop k\le i<j\le m-1} \ha_{ij}
\right)
\right)
\, .
\een
\end{enumerate}
The case where the scaling configuration involves charges $\ha_2',\cdots \ha_m'$
requires a special treatment. If we naively consider this as a special case of 
\refb{escal2} above
with $k=2$ or of \refb{escal1} for $m=4$, 
we would conclude that the jump vanishes since there are no scaling solution
with two centers and hence $G(\ha_1',\ha_2'+\cdots \ha_m')$ vanishes. However notice
that in this case \refb{especial} with $k=2$ or \refb{enewforce} for $m=4$ implies
\be   
\sum_{j=2}^m \ha'_{1j}=0
\ee
and hence the two centered configuration with one center with charge $\ha'_1$ and the other
center with charge $\ha_2'+\cdots \ha_m'$ is on a wall of threshold stability\footnote{Recall that a wall of threshold
stability is one on which the bound state can separate into two components with vanishing
DSZ product. Across this wall the topology of the bound state changes but 
 the index does not jump. 
 }. As such configurations exist
for vanishing FI parameters,
we need to analyze the situation more carefully by working at $\mu=
\mu_B+\delta\mu$  
where $\mu_B$ is the critical value of $\mu$ at which eq.\refb{especial} is
satisfied for $k=2$. At this point we have
\ben \label{enew} 
\sum_{j=2}^m \ha'_{1j} =
\begin{cases}
\delta\mu \, \ha_{1m} \quad \hbox{for $m>4$}\cr
- \delta \mu \, (\ha_{24} + \ha_{34}) \quad \hbox{for $m=4$}
\end{cases} \nonumber \\
\sum_{i,j \atop 2\le i<j \le m} \ha'_{ij} =
\begin{cases}
-\delta\mu \, \ha_{1m} \quad \hbox{for $m>4$}\cr
\delta \mu \, (\ha_{24} + \ha_{34}) \quad \hbox{for $m=4$}
\end{cases} \, .
\een
In either case we can proceed to analyze the system following a similar kind of
analysis used in computing $\Delta F_A$. We denote the locations of the centers
as $z_1$, $z_i= z_2 + y \, w_i$ for $2\le i\le m$ with $w_2\equiv 0$,  
$w_m\equiv 1$,
and look for solutions with $y\sim \delta\mu$. The solution for $z_2$ 
and $y$
are found by extremizing  
\be \label{efirst}
\wh W_1\equiv
-\sum_{i=2}^m \ha'_{1i} \, \ln(z_2-z_1) - {y \over z_2 - z_1}\sum_{i=3}^{m} \ha'_{1i} w_i
-\ln \, y \sum_{i,j\atop
2\le i<j\le m} \ha'_{ij} \, , 
\ee
with respect to $z_2$ and $y$, and  the $w_i$'s are given by extremizing
\be
\wh W_2= 
-\sum_{s,r\atop 2\le s<r\le m} \ha'_{sr} \ln (w_r-w_s)  \, ,
\ee
with respect to $w_3,\cdots w_{m-1}$. 
The extremization of $\wh W_2$ with respect to all the
$w_k$ and the sign of the
corresponding Hessian gives $G(\ha'_2,\cdots \ha'_m)$. On the other hand the
extremization with respect to $z_2$ and $y$ gives identical conditions\footnote{This 
can be traced to the fact that using scale invariance we can fix either
$z_2$ or $y$.}  
\be 
-{1\over z_2-z_1} \sum_{i=3}^{m} \ha'_{1i} w_i -{1\over y} \sum_{i,j\atop 2\le i<j\le m}
\ha'_{ij} = 0\, .
\ee
For small $\sum_{i,j\atop 2\le i<j\le m} \ta'_{ij}$, a solution for small positive $y$ exists
for only one sign of $\sum_{i,j\atop 2\le i<j\le m} \ta'_{ij}$.
The 
corresponding contribution to the sign of the Hessian can be found by taking
the second derivative of $W$ either with respect to $y$ or $z_2$ keeping the other
variable fixed, and is given by a 
multiplicative factor of  
\be
{\rm sign} \left( \sum_{ i,j \atop 2\le i<j\le m} \ha'_{ij} \right) \, .
\ee
Following a logic similar to that for $\Delta F_A$ and using \refb{enew} we find that
at this critical value, the scaling index $G$ jumps by 
\ben \label{ejumpgs}  
\Delta G_B &=&
\Theta\left(-
\left( \sum_{i,j\atop 2\le i<j\le m} \ha_{ij} \right)
\left( \sum_{i,j\atop 2\le i<j\le m-1} \ha_{ij}
+  \sum_{i=1}^{m-1} \ha_{im} \right)
\right) \nonumber \\ && \times
{\rm sign}(-\ha_{1m}) \, G(\ha'_2, \cdots \ha'_m) \quad \hbox{for $m>4$}
\nonumber \\
&=& \Theta\left(- \ha_{23}\left(\ha_{23} + \ha_{34}+\ha_{24}\right)\right)
\times {\rm sign}(\ha_{24} + \ha_{34}) \, G(\ha'_2, \cdots \ha'_4)
\quad \hbox{for $m=4$} 
\, . \nonumber \\
\een
This gives a recursive procedure for calculating the scaling index $G(\ha_1,\cdots \ha_m)$,
and therefore the total index $F$ using \refb{efinal1}, \refb{edeltafs1}.  The recursion 
is initialized by the result for three centers, given below.

\subsubsection{Coulomb index for $3$ and $4$ centers}

\begin{figure}
\setlength{\unitlength}{3947sp}%
\begin{picture}(0,0)(0,400)%
\includegraphics{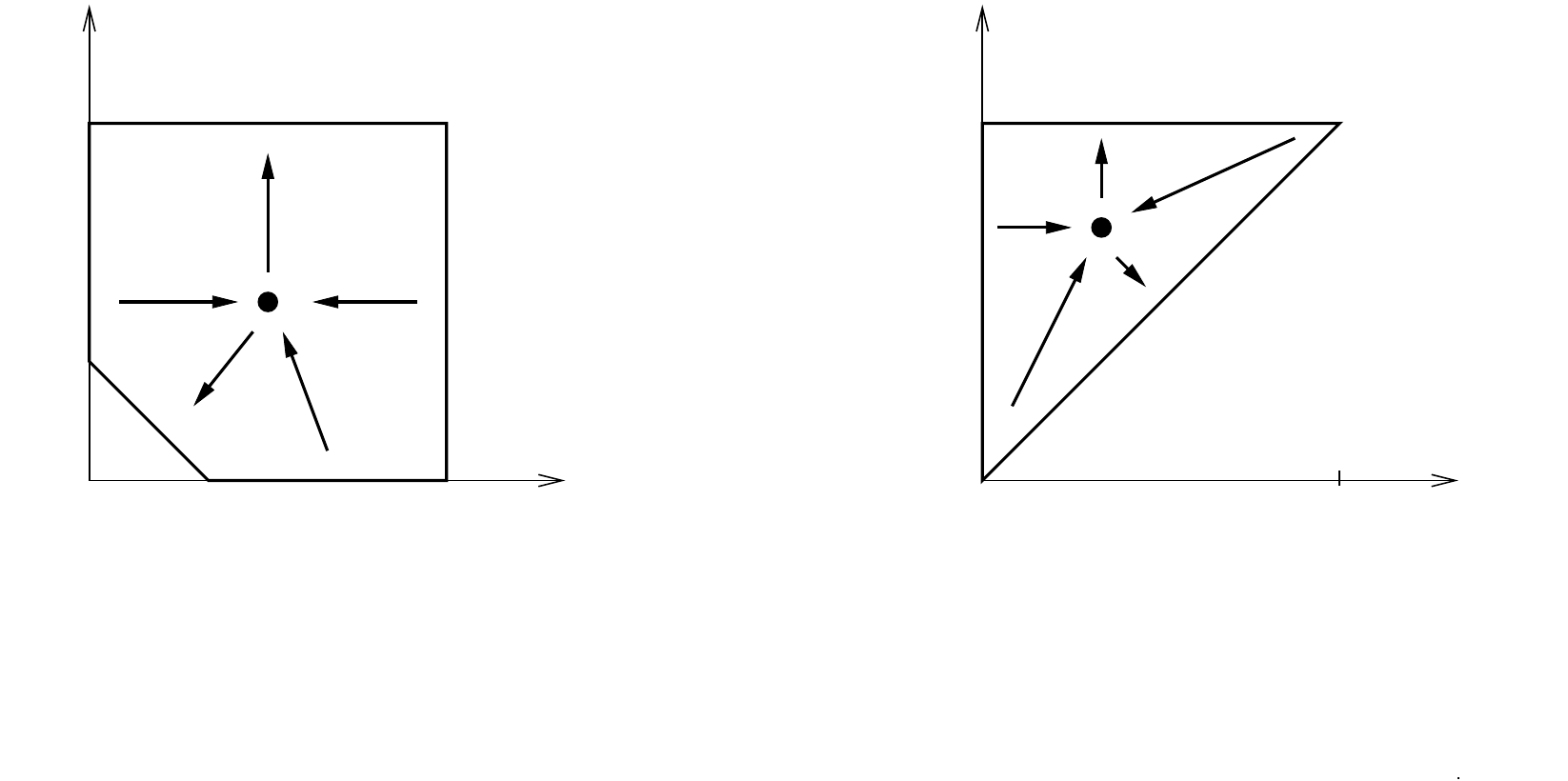}%
\end{picture}%
\begingroup\makeatletter\ifx\SetFigFont\undefined%
\gdef\SetFigFont#1#2#3#4#5{%
  \reset@font\fontsize{#1}{#2pt}%
  \fontfamily{#3}\fontseries{#4}\fontshape{#5}%
  \selectfont}%
\fi\endgroup%
\begin{picture}(7899,3924)(1351,-4773)
\put(4051,-3361){\makebox(0,0)[lb]{\smash{{\SetFigFont{12}{14.4}{\rmdefault}{\mddefault}{\updefault}{\color[rgb]{0,0,0}$y_1$}%
}}}}
\put(1951,-1411){\makebox(0,0)[lb]{\smash{{\SetFigFont{12}{14.4}{\rmdefault}{\mddefault}{\updefault}{\color[rgb]{0,0,0}$y_2$}%
}}}}
\put(1351,-2686){\makebox(0,0)[lb]{\smash{{\SetFigFont{12}{14.4}{\rmdefault}{\mddefault}{\updefault}{\color[rgb]{0,0,0}$\alpha_{12}$}%
}}}}
\put(2701,-3961){\makebox(0,0)[lb]{\smash{{\SetFigFont{12}{14.4}{\rmdefault}{\mddefault}{\updefault}{\color[rgb]{0,0,0}$\alpha_{23}$}%
}}}}
\put(3826,-2686){\makebox(0,0)[lb]{\smash{{\SetFigFont{12}{14.4}{\rmdefault}{\mddefault}{\updefault}{\color[rgb]{0,0,0}$c_1$}%
}}}}
\put(1576,-3211){\rotatebox{315.0}{\makebox(0,0)[lb]{\smash{{\SetFigFont{12}{14.4}{\rmdefault}{\mddefault}{\updefault}{\color[rgb]{0,0,0}$\alpha_{12}+\alpha_{23}+\alpha_{13}$}%
}}}}}
\put(8401,-3361){\makebox(0,0)[lb]{\smash{{\SetFigFont{12}{14.4}{\rmdefault}{\mddefault}{\updefault}{\color[rgb]{0,0,0}$z_2$}%
}}}}
\put(6451,-1411){\makebox(0,0)[lb]{\smash{{\SetFigFont{12}{14.4}{\rmdefault}{\mddefault}{\updefault}{\color[rgb]{0,0,0}$z_3$}%
}}}}
\put(8101,-3886){\makebox(0,0)[lb]{\smash{{\SetFigFont{12}{14.4}{\rmdefault}{\mddefault}{\updefault}{\color[rgb]{0,0,0}$1$}%
}}}}
\put(6001,-1936){\makebox(0,0)[lb]{\smash{{\SetFigFont{12}{14.4}{\rmdefault}{\mddefault}{\updefault}{\color[rgb]{0,0,0}$1$}%
}}}}
\put(3526,-3961){\makebox(0,0)[lb]{\smash{{\SetFigFont{12}{14.4}{\rmdefault}{\mddefault}{\updefault}{\color[rgb]{0,0,0}$\infty$}%
}}}}
\put(1501,-1861){\makebox(0,0)[lb]{\smash{{\SetFigFont{12}{14.4}{\rmdefault}{\mddefault}{\updefault}{\color[rgb]{0,0,0}inf}%
}}}}
\put(5926,-2761){\makebox(0,0)[lb]{\smash{{\SetFigFont{12}{14.4}{\rmdefault}{\mddefault}{\updefault}{\color[rgb]{0,0,0}$\alpha_{12}$}%
}}}}
\put(7051,-1711){\makebox(0,0)[lb]{\smash{{\SetFigFont{12}{14.4}{\rmdefault}{\mddefault}{\updefault}{\color[rgb]{0,0,0}$\alpha_{34}$}%
}}}}
\put(8251,-1936){\makebox(0,0)[lb]{\smash{{\SetFigFont{12}{14.4}{\rmdefault}{\mddefault}{\updefault}{\color[rgb]{0,0,0}$\alpha_{23}+\alpha_{24}+\alpha_{34}$}%
}}}}
\put(5551,-3961){\makebox(0,0)[lb]{\smash{{\SetFigFont{12}{14.4}{\rmdefault}{\mddefault}{\updefault}{\color[rgb]{0,0,0}$\alpha_{12}+\alpha_{23}+\alpha_{13}$}%
}}}}
\put(7276,-3061){\rotatebox{45.0}{\makebox(0,0)[lb]{\smash{{\SetFigFont{12}{14.4}{\rmdefault}{\mddefault}{\updefault}{\color[rgb]{0,0,0}$\alpha_{23}$}%
}}}}}
\put(2551,-1711){\makebox(0,0)[lb]{\smash{{\SetFigFont{12}{14.4}{\rmdefault}{\mddefault}{\updefault}{\color[rgb]{0,0,0}$c_1+c_2$}%
}}}}
\end{picture}
\caption{Left: The physical domain for 3-center collinear solutions has 5 boundary components
at which the superpotential $W$ diverges. The sign of $W$ on each component is that of the 
quantity indicated on the corresponding edge. Right: the physical domain for 4-center 
collinear scaling solutions also has 5 boundary components, at which the superpotential 
$\hat W$ diverges. The sign of $W$ on each component is  that of the 
linear combination of $\ha_{ij}$ indicated on the corresponding edge or vertex. In both cases, by
considering the topology of the gradient flow (indicated by the arrows for some suitable choice of signs on the boundary components), it is easy to convince oneself that
a critical point exists in the physical domain  if and only if the signs on the 5 boundary components are identical, or flip 4 times around the boundary. 
\label{figdomain}
}
\end{figure}
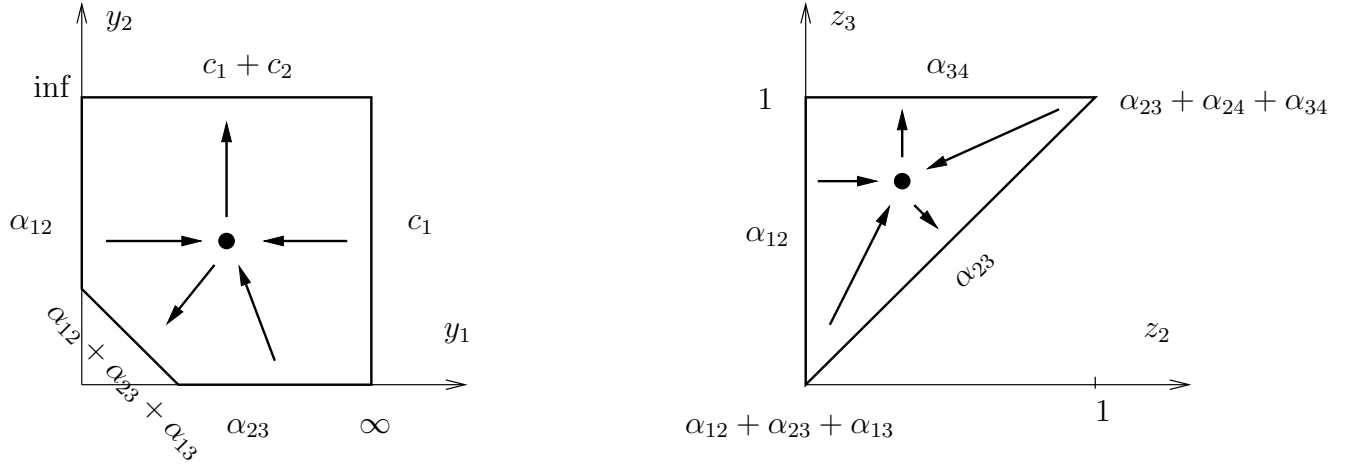

As a simple application of the procedure described above we shall calculate the Coulomb index for 3 and 4 centers. For 3 centers, collinear scaling solutions exist for
\be
{\rm sign} (\ha_{12}) = {\rm sign}(\ha_{23}), \quad \ha_{13}=
-\ha_{12}-\ha_{23}\, ,
\ee 
and the sign of the Hessian of $\widehat W$ is $(-1)^{\Theta(\ha_{23})+1}$. Thus 
\be \label{eg3pre}
G(\ha_1, \ha_2, \ha_3) = \Theta(\ha_{12} \ha_{23}) \, (-1)^{\Theta(\ha_{23})+1}\, .
\ee
The total index  given by \refb{efinal1}, \eqref{edeltafs1}, is  
\be \label{eg3}
\begin{split}
 F(\{\ta_1,\ta_2, \ta_3\}; \{\cc_1, \cc_2, \cc_3\})
= & 
(-1)^{\Theta(-\ta_{12}) +\Theta(-\ta_{23})} \, 
\Theta(\ta_{12}\, \cc_1) \, \Theta(\ta_{23} (\cc_1+\cc_2))  \\
+&  (-1)^{\Theta(\ta_{12}) + \Theta(\ta_{13})} \, \Theta(\ta_{12} \ta_{23})\, 
\Theta(-(\ta_{12}+\ta_{23}) (\ta_{12}+\ta_{23}+\ta_{13}))\, ,
\end{split}
\ee
where the first line corresponds to $F_0$ in \eqref{efinal} and the second line to 
the contribution of the scaling solution occurring at $\lambda=-(\ta_{12}+\ta_{23})/\ta_{13}$.
It is straightforward, if tedious, to check that \eqref{eg3} agrees with the result given in 
\cite{Manschot:2012rx} in a particular chamber. The result \eqref{eg3} can be succintly summarized by saying that $F$ vanishes unless the sign of the 5-periodic sequence
\be
\label{ruleF3}
\Sigma_{123} =\{\cc_1+\cc_2, \cc_1,\ta_{23},\ta_{12}+\ta_{23}+\ta_{13},\ta_{12}\}
\ee
is either constant (in which case $F=1$), or flips 4 times around the sequence (in
which case $F=-1$). These signs correspond to the 
behavior of the superpotential $W$ at the 5 boundaries $y_2=\infty, y_1=\infty, y_2=0, y_1=y_2=0,
y_1=0$ of the domain in which the variables $y_1$, $y_2$ take values. The rule
\eqref{ruleF3} is  in agreement with the existence of a gradient flow emanating from a
critical point of $W$ inside this domain (see Figure \ref{figdomain}, left).

For $n=4$, we first need to compute the scaling index $G(\ha_1,\cdots \ha_4)$.
In this case the only contribution to $\Delta G_B$ in
\refb{egrecur} comes from the configuration where the centers 2, 3 and 4 come
together during the deformation. Using \refb{ejumpgs}, \refb{eg3pre} we can
then express \refb{egrecur} as
\ben \label{eg4}
G(\ha_1,\cdots \ha_4) &=&
(-1)^{\Theta(\ha_{23}) +\Theta(\ha_{34})} \, 
\Theta(\ha_{12}\ha_{23}) \, \Theta(-\ha_{34} (\ha_{14}+\ha_{24}+\ha_{34})) 
\nonumber \\
&& + (-1)^{\Theta(\ha_{23}) + \Theta(\ha_{24}+\ha_{34})} \, \Theta(\ha_{23} \ha_{34})\, 
\Theta(-\ha_{23} (\ha_{23}+\ha_{24}+\ha_{34}))\, . \nonumber \\
\een
The rule \eqref{eg4} can be summarized
by saying that the sign of the 5-periodic sequence
\be
\Sigma_{1234} = \{ \ha_{12} , \ha_{34}, 
\ha_{23}+\ha_{34}+\ha_{24} , 
\ha_{23} , \ha_{12}+\ha_{23}+\ha_{13} \}
\ee
is either constant (in which case $G=1$) or alternates 4 times (in which case $G=-1$).
These signs correspond to the 
behavior of the superpotential $\hat W$ at the 5 boundaries $z_2=0$, $z_3=1$, $z_2=z_3=1$, $z_2=z_3$, $z_2=z_3=0$ of the domain in which the variables $z_2,z_3$ are valued (in the gauge
$z_1=0,z_4=1$). The rule
\eqref{eg4} is  in agreement with the existence of a gradient flow emanating from a
critical point of $\hat W$ inside this domain (see Figure \ref{figdomain}, right).

Using \refb{efinal1} we can now compute $F(\{\ta_1,\cdots \ta_4\},
\{\cc_1,\cdots \cc_4\}))$ as a sum
of four terms: $F_0(\{\ta_1,\cdots \ta_4\},
\{\cc_1,\cdots \cc_4\}))$ given in \refb{efinal}, and the jumps across the values of
$\mu$ where all the centers come together, where the centers 2,3,4 come together,
and where the centers 1,2,3 come together.
The final result takes the form
\be
\begin{split}
F&\left(\{\ta_1,\cdots \ta_4\},
\{\cc_1,\cdots \cc_4\}\right)  
=  
\prod_{k=1}^{3} \Theta(\ta_{k, k+1} \, \tc_k) (-1)^{\sum_{k=1}^{n-1}
\Theta(-\ta_{k,k+1})}\\
& + (-1)^{\Theta(\ta_{13}
+\ta_{14}+\ta_{24})+1}
G\left(\ta_1^{(1)}, \cdots \ta_4^{(1)}\right) 
\Theta\left( - \left(\sum_{1\le i<j\le 4}\ta_{ij}\right) (\ta_{12}+\ta_{23}+\ta_{34})\right)
 \\
& + (-1)^{\Theta(\ta_{24})+1} F\left(\{\ta_1^{(2)}, \ta_2^{(2)}+\ta_3^{(2)}+\ta_4^{(2)}\},
\{\cc_1, \cc_2+\cc_3+\cc_4\}\right) G\left(\ta_2^{(2)},\ta_3^{(2)},\ta_4^{(2)}\right)
 \\
& \qquad \times \Theta(-(\ta_{23}+\ta_{34})(\ta_{23}+\ta_{34}+\ta_{24}))
\\
& +(-1)^{\Theta(\ta_{13})+1} F\left(\{ \ta_1^{(3)}+\ta_2^{(3)}+\ta_3^{(3)}, \ta_4^{(3)}\},
\{ \cc_1+\cc_2+\cc_3, \cc_4\}\right) G\left(\ta_1^{(3)},\ta_2^{(3)},\ta_3^{(3)}\right) 
\\
& \qquad \times \Theta\left( - (\ta_{12}+\ta_{23}) (\ta_{12}+\ta_{23}+\ta_{13})\right)\, ,
\end{split}
\ee
where
\ben
&&\ta^{(1)}_{ij} = \lambda_1 \, \ta_{ij} \quad \hbox{for $|i-j|\le 2, \, 1\le i,j\le 4$},
\quad \ta^{(1)}_{ij} = \ta_{ij} \quad \hbox{otherwise}\, , \nonumber \\
&&\ta^{(2)}_{24} = \lambda_2 \, \ta_{24} , \quad 
\ta^{(2)}_{ij} = \ta_{ij} \quad \hbox{otherwise}\, , \nonumber \\
&&\ta^{(3)}_{13} = \lambda_3 \, \ta_{13} , \quad 
\ta^{(3)}_{ij} = \ta_{ij} \quad \hbox{otherwise}\, ,
\een
\be
\lambda_1= -{\ta_{12}+\ta_{23}+\ta_{34}\over \ta_{14}+\ta_{24}+\ta_{13}},
\quad \lambda_2 = -{\ta_{23}+\ta_{34}\over \ta_{24}}, \quad
\lambda_3 = - {\ta_{12}+\ta_{23}\over \ta_{13}}\, .
\quad 
\ee
This can  be easily generalized to higher number of centers.

\section{Quiver 
invariants} \label{sec_nonAbelian}

In this section we shall describe how the results of the previous sections can be
used to give a complete prescription for computing the Poincar\'e-Laurent polynomial of quiver
moduli spaces. For this we need to briefly review the prescription given in
\cite{Manschot:2012rx}.

\subsection{Quiver Poincar\'e-Laurent polynomial from
Coulomb index: a review}
\label{sec_Poincare}

We shall
consider a quiver with $K$ nodes with a $U(N_\ell)$ factor at the $\ell$-th node,
$\gamma_{\ell\kk}$ arrows from the $\ell$-th node to the $\kk$-th node
representing $\gamma_{\ell\kk}$ number of $(N_\ell, \bar N_k)$ representations
of $U(N_\ell)\times U(N_{\kk})$ and FI parameters $\zeta_1,\cdots\zeta_K$
satisfying $\sum_\ell N_\ell \zeta_\ell=0$. 
A negative $\gamma_{\ell\kk}$ indicates $\gamma_{\kk\ell}\equiv -\gamma_{\ell\kk}$ 
number of $(\bar N_\ell, N_{\kk})$ representations of
$U(N_\ell)\times U(N_{\kk})$.
Instead of considering one specific quiver at a time it turns out to be more
convenient to consider the  family of quivers labelled by different ranks $\{N_\ell\}$
and different values of FI parameters $\{\zeta_\ell\}$.
For this we 
assign to each node $\ell$ a basis vector $\gamma_\ell=(0,\dots, 0,1,0,\dots 0)$ --
with 1 inserted at the $\ell$-th position -- in an
abstract vector space $\IZ^K$, denote by 
$\Gamma\subset \IZ^K$ the collection of vectors
$\gamma=\sum_{\ell=1}^K N_\ell \gamma_\ell$ where $N_\ell$
are non-negative integers, and by $\cC_\gamma$ 
the hyperplane $\sum_{\ell=1}^K N_\ell \zeta_\ell=0$
in the space of real vectors $\zeta=\sum_{\ell=1}^K \zeta_\ell \gamma_\ell \in \IR^K$. 
We also
introduce a 
symplectic inner product
\be
\label{defEulerForm}
\langle \gamma, \gamma'\rangle\equiv\sum_{\ell, \kk=1 }^K 
N_\ell N'_\kk\,  \gamma_{\ell\kk},
\ee
between the elements $\gamma=\sum_{\ell=1}^K N_\ell \gamma_\ell$ of $\Gamma$.
To any vector $\gamma\in\Gamma$ and $\zeta\in\cC_\gamma$, 
we associate a quiver $\cQ(\gamma,\zeta)$
with $K$ nodes, $\gamma_{\ell\kk}$
arrows connecting the node $\ell$ to the node $\kk$, gauge group
 $U(N_1)\times U(N_2)\times \cdots U(N_K)$, and FI parameters $\{\zeta_1,\cdots \zeta_K\}$.
If some of the $N_\ell$'s vanish we just drop the corresponding nodes.

Let $Q(\gamma; \zeta;y)$ be the Poincar\'e-Laurent polynomial
\be \label{epol}  
Q(\gamma; \zeta;y)  = \sum_{p=1}^{2d} b_p(\cM)\, (-y)^{p-d}
\ee
where $d$ is the complex 
dimension of the moduli space $\cM$ of the quiver 
$\cQ(\gamma;\zeta)$
and 
the $b_p$'s are the topological Betti numbers of $\cM$.
The Coulomb branch formula  
for $Q(\gamma; \zeta;y)$,  
which we denote by $\QC(\gamma; \zeta;y)$, takes the form:
\ben \label{essp1}
\QC(\gamma; \zeta;y) &=& \sum_{m|\gamma} 
\frac{\mu(m)}{ m}  {y - y^{-1}\over y^m - y^{-m}}
\bQC(\gamma/m; \zeta;y^m) \nonumber \\
\bQC(\gamma; \zeta;y) &=& 
\sum_{n\ge 1}\sum_{\{\alpha_i\in \Gamma\} \atop \sum_{i=1}^n \alpha_i =\gamma}
 \frac{\gref\left(\{\alpha_1, \cdots, \alpha_n\},
 \{\dx_1,\cdots \dx_n\};y\right)}
{  |{\rm Aut}(\{\alpha_1,\cdots, \alpha_n\})|}
  \nonumber \\ &&\quad
\prod_{i=1}^n \Big( \sum_{m_i\in\bZ\atop m_i|\alpha_i}
{1\over m_i} {y - y^{-1}\over y^{m_i} - y^{-m_i}}\, 
\Omega_{\rm tot}(\alpha_i/m_i;y^{m_i})
\Big)
\, .
\een
The first line is the standard relation between integer and rational BPS invariants
which has appeared in a variety of contexts \cite{Chuang:2010ii, 1011.1258, Manschot:2011ym, 1107.0723}.  $\mu(m)$ is the M\"obius function, which is 1 ($-1$) if $m$ is a square-free positive integer with an even (odd) number of prime factors, and 0 if $m$ is not square-free. 
In the second line,
$|{\rm Aut}(\{\alpha_1,\cdots \alpha_n\})|$ is a symmetry factor given by
$\prod_k s_k!$ if among the set $\{\alpha_i\}$ there are
$s_1$ identical vectors $\tilde \alpha_1$, $s_2$ identical vectors
$\tilde\alpha_2$ etc., and $m|\alpha$ means that $m$ is a common divisor of
$(n_1,\cdots , n_K)$ if $\alpha =\sum_\ell n_\ell \gamma_\ell$.
The sums over $n$ and $\{\alpha_1,\cdots \alpha_n\}$ in the second
equation label all possible ways of
expressing $\gamma$ as (unordered) sums of elements $\alpha_i$ of $\Gamma$. 
The coefficients
$\dx_i$ are determined in terms of the FI parameters $\zeta_i$ by $\dx_i
=\sum_\ell A_{i\ell} \zeta_\ell$ whenever  $\alpha_i=\sum_\ell A_{i\ell}\gamma_\ell$. 
From the restrictions $\sum_i \alpha_i
=\gamma$ and $\sum_\ell N_\ell \zeta_\ell=0$ it follows that 
$\sum_i \dx_i=0$.
The Coulomb indices $\gref(\{\alpha_1,\cdots, \alpha_n\};
 \{\dx_1,\cdots \dx_n\};y)$ can be computed from \refb{eqforg}.
The functions $\Omega_{\rm tot}(\alpha;y)$ 
are expressed
in terms of the single-centered BPS invariants $\OmS$ through
\be \label{essp2}
\Omega_{\rm tot}(\alpha;y) = \OmS(\alpha;y) + 
\sum_{\{\beta_i\in \Gamma\}, \{m_i\in\bZ\}\atop
m_i\ge 1, \, \sum_i m_i\beta_i =\alpha}
H(\{\beta_i\}; \{m_i\};y) \, \prod_i 
\OmS(\beta_i;y^{m_i})
\, .
\ee
Finally, the functions $H(\{\beta_i\}; \{k_i\};y)$ and $\OmS(\gamma;y)$
are
determined as follows. 
\begin{enumerate}
\item When the number of
$\beta_i$'s is less that three,  $H(\{\beta_i\}; \{k_i\};y)$ vanishes.
\item
For three or more number of $\beta_i$'s, 
observe that  the expression for $\QC(\sum_i k_i\beta_i; \zeta;y)$ given in
\eqref{essp1} contains a term proportional to
$H(\{\beta_i\}; \{k_i\};y)\prod_i\OmS(\beta_i;y^{k_i})$
arising from the choice $m=1$ in the first equation in
\eqref{essp1}, $n=1$,
$\alpha_1=\sum_i k_i\beta_i$, $m_1=1$
in the second equation in \eqref{essp1}, and 
$m_i=k_i$ in the expression for
$\Omega_{\rm tot} (\sum_i k_i \beta_i; y)$ in eq.\eqref{essp2}.
We fix $H(\{\beta_i\}; \{k_i\};y)$ by demanding that the net
coefficient of the product $\prod_i \OmS(\beta_i;y^{k_i})$ in the
expression for $\QC(\sum_i k_i\beta_i; y)$ is a
Laurent polynomial in $y$. This of course leaves open the possibility  of
adding to $H$ a
Laurent polynomial. This is resolved by using the {\it minimal
modification hypothesis}, which requires that $H$ must 
be symmetric under $y\to y^{-1}$ and vanish
as $y\to\infty$  \cite{1103.1887}.
We determine $H(\{\beta_i\}; \{m_i\};y)$ iteratively by
beginning with the $H$'s with three $\beta_i$'s and then 
determining successively the $H$'s with more $\beta_i$'s. 
\item
$H$ is expected 
to be independent of the
FI parameters and hence can be calculated for any
value of these parameters.
\item
{\it After determining $H(\{\beta_i\};\{k_i\};y)$ in this way}, we set 
$\OmS(\gamma_\ell;y)=1$ for $1\le \ell\le K$. For
all other charge vectors $\beta$,  $\OmS(\beta;y)$
are fixed integers, independent of $y$ and of the
FI parameters, which are left undetermined by the Coulomb branch analysis. 
Since these unknown constants, as well as the quivers, are 
labelled by the vectors $\alpha\in \Gamma$, there is one\footnote{Actually the 
number of unknown constants is less than that
of the number of quivers since $\OmS(\gamma)$ is non-trivial only 
if there exists a set of $\alpha_i$'s in $\Gamma$ 
such that $\sum_i\alpha_i=\gamma$ and it is possible to find three dimensional
vectors $\vec r_i$
such that $\sum_{i,j} \alpha_{ij} (\vec r_i - \vec r_j) / |\vec r_i-\vec r_j|=0$.}
 unknown
constant for each quiver. This can be fixed e.g. by computing the Euler character of the
quiver moduli space for any convenient value of the FI parameters. 
\end{enumerate}

As a special case of our result we can consider the case of a general Abelian
quiver. This corresponds to $\gamma=\sum_\ell N_\ell \gamma_\ell$ 
with $N_\ell=0$ or 1. As a result $\gamma$, as well as the $\alpha_i$'s
appearing on the right hand side of \refb{essp1}, are primitive vectors and
the $\alpha_i$'s are all distinct. Thus
\refb{essp1}, \refb{essp2} simplifies to
\ben \label{essp1spec}
\QC(\gamma; \zeta;y) &=&
\sum_{n\ge 1}\sum_{\{\alpha_i\in \Gamma\}\atop \sum_{i=1}^n \alpha_i =\gamma}
 \gref\left(\{\alpha_1, \cdots, \alpha_n\},
 \{\dx_1,\cdots \dx_n\};y\right)
\prod_{i=1}^n
\Omega_{\rm tot}(\alpha_i;y)
\, ,
\een
\be \label{essp2spec}
\Omega_{\rm tot}(\alpha;y) =\OmS(\alpha;y) + 
\sum_{\{\beta_i\in \Gamma\}, \sum_i \beta_i =\alpha}
H(\{\beta_i\};y) \, \prod_i 
\OmS(\beta_i;y)
\, ,
\ee
where $H(\{\beta_i\};y)\equiv H(\{\beta_i\};\{1,1,\cdots 1\};y)$. 
The functions $H(\{\beta_i\};y)$ are determined by requiring that they vanish as
$y\to 0, \infty$, are invariant under $y\to y^{-1}$, and that the coefficient of
$\prod_i 
\OmS(\alpha_i;y)$ in the expression for $\QC(\gamma; \zeta;y)$ is
a positive integer for each set $\{\alpha_i\}$.

\subsection{Coulomb index for non-generic charges} \label{sec_nongeneric}

The formul\ae\ \refb{essp1}, \refb{essp2}
are completely explicit provided we have
an explicit  algorithm for
computing $\gref(\{\alpha_1,\cdots \alpha_n\};\{\dx_1,\cdots \dx_n\};y)$. 
We have given such an algorithm
in the previous sections for generic $\alpha_{ij}$'s, {\it e.g.} all 
$\alpha_{ij}$'s non-zero and no ordered subset $\bar\alpha_1,\cdots \bar\alpha_s$
of the $\alpha_i$'s satisfying $\sum_{1\le k<\ell\le s}\bar\alpha_{k\ell}=0$.
We have also assumed that the FI parameters stay away from the walls of marginal and
threshold stability so that {\it e.g.} the quantities $\sum_{i=1}^k c_{\sigma(i)}$
appearing in \refb{eindexfin} never vanish.
However we need $\gref$ for
non-generic $\alpha_{ij}$'s and $c_i$'s as well. 
These come from two sources. First of all the
$\gamma_{\ell k}$'s of the original quiver themselves may be non-generic with some
$\gamma_{\ell k}$'s vanishing or satisfying special relations. Second, even if the
original $\gamma_{\ell k}$'s are generic, in the argument of $\gref$ we may
have parallel  $\alpha_i$'s. For these the corresponding $\alpha_{ij}$'s
will vanish. 
Also when the total dimension vector $\{N_1,\cdots N_K\}$
is non-primitive, we shall encounter $\gref$ in \refb{essp1} for which the FI parameters sit
on the threshold stability walls.
In all such cases we need to evaluate $\gref$ by first deforming the
$\alpha_{ij}$'s and/or $c_i$'s 
to generic values and then taking the limit back to the original
configuration. The goal of this section will be to determine a prescription for
such deformations.
We shall first describe the prescription  and then justify it.

\begin{enumerate}
\item
To deal with the first problem we deform the $\gamma_{\ell k}$'s to
\be
\label{deform1}
\gamma_{\ell k}\to \gamma_{\ell k} + \epsilon_1 \, \xi_{\ell k}
\ee
where $\epsilon_1$ is a small positive
number and $\xi_{\ell k}$'s are random numbers
between $-1$ and $1$ satisfying $\xi_{\ell k}=-\xi_{k\ell}$. 
This will make all the $\gamma_{\ell k}$ generic.\footnote{In order to
increase the efficiency of the procedure, we can arrange the nodes in
some fixed order and then choose the $\xi_{\ell k}$'s such that
$\xi_{\ell k}>0$ for $\ell<k$. This will minimize the introduction of
new oriented loops and hence
scaling configurations during this deformation. For example if we
have three nodes $i$, $j$ and $k$ such that $\alpha_{ij}=\alpha_{jk}=\alpha_{ik}=0$,
then under the deformation \refb{deform1} the subquiver containing the
nodes $i$, $j$ and $k$ will not have any oriented loop.}
\item At this stage in any given term in the sum in \refb{essp1},
all the $\alpha_{ij}$'s are generic except for subsets of $\alpha_i$'s
which are all parallel and/or equal. In computing
$\gref(\{\alpha_i\},\{\dx_i\};y)$ for a set of $\alpha_i$'s we 
consider an arbitrary ordering\footnote{Here we are considering the
ordering as a set and not an ordering of the locations of the centers.
The same deformation must be used for all possible arrangements of the
centers.
} 
of all
the $\alpha_i$'s and deform them by
\be 
\label{deform2}
\alpha_{ij} \to \alpha_{ij} + \epsilon_2 \, \beta_{ij}, \qquad c_i\to c_i +\epsilon_2 f_i\, ,
\ee
where $\epsilon_2$ is a small positive
number that is parametrically smaller than the
previous paramater $\epsilon_1$, $\beta_{ij}$'s for $i<j$ are
randomly chosen {\it positive}
numbers between 0 and 1 with $\beta_{ji}=-\beta_{ij}$, and $f_i$'s are random numbers
satisfying $\sum_i f_i=0$. Under such a deformation any subset of the $\{\alpha_i\}$'s
which are parallel and/or equal to each other get deformed in such a way that the 
corresponding subquiver does not contain any oriented loop. Also the FI parameters
move away from threshold stability walls even if the undeformed configuration sits on
such a wall.
\item
At the end of the second step the $\alpha_{ij}$'s and $c_i$'s
are generic and can be used to
compute $\gref$.
In particular the $s(\sigma)$ factors in $\gref$ are computed using the
deformed $\alpha_{ij}$'s and $c_i$'s. However in computing the $y^{\sum_{i<j}\ta_{ij}}$ factors
in $\gref$ we use the undeformed $\alpha_{ij}$'s, since at the end of the
computation we are in any case supposed to take the $\alpha_{ij}$'s to their
undeformed values.
This $\gref$ is then used to compute $\QC(\gamma; \zeta;y)$ via \refb{essp1}.
\end{enumerate}
 
 In order to prove the validity of the procedure we need to argue that 
the deformed result reduces to the undeformed one  in the limit when the 
deformations are switched off. 
First note that there is a qualitative difference between the
 deformations generated by $\epsilon_1$ and those generated by $\epsilon_2$. For the
 latter the deformation of the $\alpha_{ij}$'s and $c_i$'s
 we use is specific to the $\alpha_i$'s and $c_i$'s which
 appear in the argument of a given $\gref$. As a result we need to establish that
 each $\gref$ returns to its undeformed value upon switching off this 
 deformation. On the other hand, the deformation generated by $\epsilon_1$ can be
 carried out for the full index $\QC(\gamma;\zeta;y)$
 as it applies to the whole family of quivers
 and does not refer to any specific multi-centered black hole configuration.
 We shall indeed argue that while the
 individual $\gref$'s in the $\epsilon_1$ deformed system do not necessarily 
 reduce to the
 undeformed result, the total index does. 
 
 Let us begin with the $\epsilon_2$ deformation. 
 This contains two parts: deformation of the $\alpha_{ij}$'s and deformation of
 the $c_i$'s. First consider the effect of deforming the $c_i$'s. If the initial
 configuration is away from the walls of marginal and threshold stability then this
 deformation has no effect. However when $\gamma=\sum_i\alpha_i$ is not primitve,
 it could happen that the set $\{\alpha_1,\cdots \alpha_n\}$ can be divided into
 two or more sets such that the sum of the $\alpha_i$'s in each set is parallel to
 the total charge $\gamma$. In this case the sum of the $c_i$'s in each set
 vanishes and the FI parameters sit on the wall of threshold stability. The deformation of the
 $c_i$'s given in \refb{deform2} is needed to move away from this wall and make $\gref$
 well defined, but the result does not depend on how we deform the $c_i$'s.

 Next we turn the effect of the $\epsilon_2$ deformation on the $\alpha_{ij}$'s.
 We begin with the deformed system and take the $\alpha_{ij}$'s one by one back to
 their values after the first deformation. During this process $\gref$ can jump if
 two or more centers come together during the deformation.
 Using the analysis of \S\ref{sec_noloopany} we know that the 
 possible jumps in $\gref$ could arise if during the deformation
 a subset $A$ of the centers can come in the collinear scaling configuration by having
 $\sum_{i<j; i,j\in A} \alpha'_{ij}=0$. Since 
 at this stage we have already
 carried out the $\epsilon_1$ deformation making the $\gamma_{\ell\kk}$'s generic;
 the possible subsets where this could happen will only involve the centers carrying
 equal or parallel charges at the end of the first deformation. However this is ruled out
 by the fact that the second deformation has been choosen so that any subquiver,
 containing equal or parallel charges at the end of the first deformation, 
 remains free from oriented loops. This
 shows that we do not encounter any collinear scaling solutions during the second
 deformation and hence there is no jump in the index $\gref$ during this
 deformation.
 
Finally we  turn to the $\epsilon_1$ deformation. 
To deal with this case we note that the analysis of \S\ref{sec_noloopany}, showing that the
refined index of a quiver changes continuously under deformations of
the $\alpha_{ij}$'s, breaks down     
on a subspace on which  $\sum_{i,j\in A; i<j} \ta_{ij}=0$ for some
subset $A$. Around this subspace the Coulomb index computed for a given set of charges
will depend on the sign of the deformation parameters since a particular collinear
solution may exist for one sign of the deformation, but as the deformation parameter
approaches zero the centers in the subset $A$ come close together and for the
opposite sign of the deformation parameter the solution ceases to exist.
However we shall now argue that $\QC(\gamma; \zeta;y)$ computed from \refb{essp2}
remains independent of the sign of the deformation. For this suppose  that 
we have three centers carrying charges $\alpha_1$, $\alpha_2$ and $\alpha_3$
such that $\alpha_{12}+\alpha_{23}+\alpha_{13}=0$, 
$\alpha_{12}, \alpha_{23}>0$. In that case for 
the permutation $(123)$ we can have a collinear scaling solution. Now in the deformed
system whether this permutation contributes or not depends on whether
$\alpha_{12}$, $\alpha_{23}$ and $\alpha_{31}$ form an oriented triangle or not, and
this in turn will depend on the details of the deformation. The difference in 
$\gref(\{\alpha_1,\alpha_2,\alpha_3\}; \{c_1,c_2,c_3\};y)$ 
that we shall get between these two cases is, up to a sign, given by
$(y-y^{-1})^{-2}$ since $y^{\sum_{i<j}\alpha_{ij}}=1$. 
In the expression for $\QC(\alpha_1+\alpha_2+\alpha_3; \zeta;y)$
there will be a term proportional to 
$\gref(\{\alpha_1,\alpha_2,\alpha_3\};\{\dx_1,\dx_2,\dx_3\};y)
\OmS(\alpha_1)\OmS(\alpha_2)
\OmS(\alpha_3)$ and the ambiguity in $\gref$ described above will
lead to an ambiguity proportional to 
$(y-y^{-1})^{-2} \OmS(\alpha_1)\OmS(\alpha_2)
\OmS(\alpha_3)$ in the expression for
$\QC(\alpha_1+\alpha_2+\alpha_3; \zeta;y)$. Now the expression for 
$\QC(\alpha_1+\alpha_2+\alpha_3; \zeta;y)$ will also contain a term proportional to
$H(\{\alpha_1,\alpha_2,\alpha_3\};y) \OmS(\alpha_1) 
\OmS(\alpha_2)
\OmS(\alpha_3)$, and the function
$H(\{\alpha_1,\alpha_2,\alpha_3\};y)$ is determined by requiring that
$H$ vanishes as $y\to 0,\infty$ and that the net coefficient of
$\OmS(\alpha_1)\OmS(\alpha_2)
\OmS(\alpha_3)$ is a polynomial in $y,y^{-1}$. Since
$(y-y^{-1})^{-2}\to 0$ as $y\to 0,\infty$ we see that the ambiguity in $\gref$
introduced above is absorbed completely into the function $H$ and does not
lead to any ambiguity in the expression for $\QC(\alpha_1,\alpha_2,\alpha_3; \zeta;y)$.
This argument can be easily generalized to argue that all the ambiguities
in $\gref$ can be absorbed into the functions $H$ and the expression for
$\QC(\gamma; \zeta;y)$ is independent of the choice of
the deformation of the $\gamma_{\ell \kk}$'s for general vector $\gamma$.

Using the analysis above we can also answer the question: 
how small should $\epsilon_1$ and $\epsilon_2$ 
 be for the procedure described above to hold? The general rule is that if 
 $\alpha'_{ij}$ denotes the deformed $\alpha_{ij}$, then during the
 deformation we should not have $\sum_{i<j;i,j\in A} \alpha'_{ij}=0$ for any ordered
 subset
 $A$ of the centers. This means that we should not encounter
 any collinear scaling configurations during the deformation except possibly at the
 beginning.\footnote{This would happen if the initial configuration had
 $\sum_{i<j; i,j\in A}\alpha_{ij} =0$ for some ordered subset $A$ of the centers.}

In \cite{Manschot:2012rx} we also proposed a formula for the Dolbeault polynomial
\be \label{edolbeault}
Q(\gamma; \zeta; y,t) \equiv \sum_{p,q} h_{p,q} (-y)^{p+q-d} t^{p-q}\, ,
\ee 
where $h_{p,q}$ are the Hodge numbers of $\cM$. The formula took the same
form as \eqref{essp1}, \eqref{essp2}, with the only difference that $\Omega_S$ was
allowed to depend on $t$, and in \eqref{essp1}, \eqref{essp2}, all factors of
$\OmS(\alpha;y^m)$ were replaced by $\OmS(\alpha;y^m;t^m)$. Eventually
we drop the $y$-dependence of $\OmS$, but they continue to depend on $t$,
giving a $t$-dependent formula for $Q(\gamma; \zeta; y,t)$.

\sectiono{Coulomb/Higgs equivalence for quivers
without loops} \label{sec_equiv} 

For quivers without oriented loops, quiver invariants can be computed using the Harder-Narasimhan recursion, or equivalently using Reineke's solution to this
recursion \cite{MR1974891}. 
In this section we shall show that for such quivers the Coulomb branch formula \refb{essp1} 
agrees with Reineke's formula, both for Abelian and non-Abelian quivers.

\subsection{Reineke's formula for quivers  without loops}

As reviewed in \cite{Manschot:2012rx}, the Poincar\'e-Laurent polynomial of a quiver 
$\cQ$ without oriented loops can be computed using the 
Harder-Narasimhan recursion.
Henceforth we shall denote the  Poincar\'e-Laurent polynomial
computed by this method by $Q_{\rm Higgs}
  (\sum_\ell N_\ell \gamma_\ell;\zeta;y)$, to distinguish it from the Coulomb branch
  formula \refb{essp1}.
The expression for $\QR$ takes the form \cite{0811.2435},\cite[Theorem 6.8]{Joyce:2004} 
\be
\label{eq:inversestackinv}
\begin{split}
\QR(\gamma; \zeta;y) =& \sum_{m|\gamma} 
\frac{\mu(m)}{ m}  {y - y^{-1}\over y^m - y^{-m}}
\bar Q_{\rm Higgs} (\gamma/m; \zeta;y^m) \\
\bar Q_{\rm Higgs}
  \left(\sum_\ell N_\ell \gamma_\ell;\zeta;y\right) 
  =& \sum_\ell 
  \sum_{\{\vec M^{(i)}\} \atop
  \sum_{i=1}^\ell \vec M^{(i)}=\vec N; \vec M^{(i)}\parallel \vec N
  \,\mathrm{for}\,\, i=1,\dots,\ell}
\frac{1}{\ell\,(y-1/y)^{\ell-1}}  \\
& \times \prod_{i=1}^\ell 
\gR(\{ M^{(i)}_1, \cdots M^{(i)}_K \}; \{\gamma_1,\cdots \gamma_K\}; 
\{\zeta_1, \cdots \zeta_K\};y)
\, .  
\end{split}
\ee
The sum over $\{\vec M^{(i)}\}$ runs over all {\it ordered} partitions of
 $\vec N\equiv (N_1,\cdots N_K)$ into parallel vectors $\vec M^{(1)}, \cdots \vec M^{(\ell)}$, and
  $\gR(\{ M^{(i)}_1, \cdots M^{(i)}_K \}; \{\gamma_1,\cdots \gamma_K\}; 
\{\zeta_1, \cdots \zeta_K\};y)$ is the `stack invariant' associated to the quiver 
with dimension
  vector $\vec M^{(i)}$. If $\vec M^{(i)}$ is primitive then
  the quiver moduli space is smooth and $\gR$ is just its Poincar\'e-Laurent polynomial, however 
  in general it is a rational function of $y$, not necessarily invariant under $y\to 1/y$. 
  In all cases however, we assume that it is given by Reineke's 
  formula \cite{MR1974891},\footnote{It
is related to the quantity $\mathcal{I}(\gamma,w)$ defined in 
\cite{Manschot:2012rx}, Eq. (2.39) via 
$\gR(\gamma;\zeta;y)=(1/y-y) \, \mathcal{I}(\gamma;-y)$.}
\be \label{ec5}
\begin{split}
 \gR& (\{M_1,\cdots M_L\};
\{\alpha_1,\cdots \alpha_L\}; \{c_1, \cdots c_L\};y) 
=  (-y)^{-  \sum_{i,j} M_i M_j \max(\alpha_{ij},0) -1 +\sum_i M_i} \,
\\  & \times  (y^2-1)^{1-\sum_i M_i} \, \sum_{\rm partitions}  (-1)^{s-1} y^{2\sum_{a\leq b}\sum_{i,j}\max(\alpha_{ij},0) \, 
N^b_i N^a_j} \prod_{a,i} ([N^a_i,y]!)^{-1}\, ,
\end{split}
\end{equation}
where 
the sum over partitions in \eqref{ec5}
runs over all {\it ordered} partitions of 
the vector $(M_1,\cdots M_L)$ into 
non-zero vectors $\{
(N^a_1,\cdots N^a_L), \quad a=1,\dots, s\}$
for $s=1,\dots, \sum_i {M_i}$,
satisfying $N^a_i\ge 0$, 
$\sum_a N^a_i=M_i$ and (assuming that $\sum_i M_i c_i=0$)
\be \label{econdombetana} 
\sum_{a=1}^b \, \sum_{i=1}^L N^a_i  c_i > 0 
\ee 
for all $b$ between 1 and $s-1$.
$[N,y]!$ denotes the $q$-deformed factorial,
\be \label{ec5.5}
[N,y]! \equiv
[1,y][2,y]\ldots[N,y]\, ,\qquad 
[N,y] \equiv \frac{y^{2N}-1}{y^2-1}\ .
\ee
It is worth noting that \refb{eq:inversestackinv} can be inverted to express  
the stack invariants in terms of the 
rational Poincar\'e-Laurent polynomials\cite[Theorem 6.8]{Joyce:2004}:
\be
\label{gtoQb}
\begin{split}
 \gR&(\{N_1,\cdots N_K\}; \{\gamma_1,\cdots \gamma_K\};
\{\zeta_1,\cdots \zeta_K\};y)  \\
=& \sum_k \sum_{\substack{\{\alpha_i\} \\
\sum_{i=1}^k \alpha_i=\sum_\ell N_\ell \gamma_\ell\\
\alpha_i\parallel \sum_{N_\ell \gamma_\ell}\,\mathrm{for}\,\, i=1,\dots, k} }
\, \frac{(-1)^{k-1}}{k! (y-y^{-1})^{k-1}}
\prod_{i=1}^k\, \bar Q_{\rm Higgs}(\alpha_i;\zeta;y)\, ,
\end{split}
\ee
where again the sum over $\{\alpha_i\}$ runs over all {\it ordered} partition of
 $\sum_\ell N_\ell \gamma_\ell$ 
 into parallel vectors $\alpha_1,\alpha_2,\cdots \alpha_k$.
 We shall refer to the formula \refb{eq:inversestackinv} (or equivalently
 \refb{gtoQb}) and
\refb{ec5} 
as the Higgs branch formula for the Poincar\'e-Laurent polynomial $Q(\gamma;\zeta;y)$.
Our goal will be to show the equality of this Higgs branch result  
with the Coulomb branch formula \eqref{essp1}.

If $\sum_\ell N_\ell \gamma_\ell$ is a primitive vector, \i.e.\ the $N_\ell$'s have
no common factor other than unity, then \refb{eq:inversestackinv} reduces to
a simple form
\begin{equation} 
\label{ec5a}
\QR\left(\sum_{i=1}^K N_i \gamma_i; \zeta;y\right) = 
\gR(\{N_1,\cdots N_K\};
\{\gamma_1,\cdots \gamma_K\}; \{\zeta_1, \cdots \zeta_K\};y)\, .
\ee
In this case $\gR$ is a symmetric Laurent polynomial since $\QR$ is.

\subsection{Abelian quivers without loops} \label{sre1}  \label{sec_equivAbelian}

For  quivers with dimension vector $N_i=1$, the stack
invariant $\gR$ appearing in \refb{ec5a} 
is given by \eqref{ec5} where
the integers
$N_i^a$  can only be equal to 0 or 1. We shall use a special symbol
$\gRa$ for labelling the corresponding $\gR$:
\be \label{edefghiggs}
\gRa(\{\gamma_1,\cdots \gamma_K\}; \{\zeta_1, \cdots \zeta_K\};y)
\equiv \gR(\{1,\cdots 1\};
\{\gamma_1,\cdots \gamma_K\}; \{\zeta_1, \cdots \zeta_K\};y)\, .
\ee
Assuming for the moment that  all $\gamma_{ij}$'s are non-vanishing, 
we shall choose a strict ordering convention for the $\gamma_i$'s such
that $\gamma_{ij}>0$ iff $i<j$. Thus \refb{ec5a}, \refb{ec5}
can be expressed as
\be
\label{thebigformula}
\begin{split}
\QR & (\gamma_1+\cdots+\gamma_K;\zeta;y) =\gRa(\{\gamma_1,\cdots \gamma_K\}; 
\{\zeta_1, \cdots \zeta_K\};y) = \\
& (-1)^{-K+1+\sum_{i<j}\gamma_{ij}} \,
 (y-y^{-1})^{1-K} \, 
 \sum_{\rm partitions}(-1)^{s-1} y^{2\sum_{a\leq b}\sum_{j<i }
 \gamma_{ji}\, 
 N_i^a \, N_j^b  - \sum_{1\le i<j\le K} \gamma_{ij}}\, . 
\end{split}
\ee
where the sum runs over all ordered partitions of 
$\gamma=\gamma_1+\cdots + \gamma_K$ into vectors $\beta^{(a)}=\sum_i N_i \gamma_i$ with
$N_i^a=0,1$, $a=1,\cdots s$, satisfying
\be
 \label{econdombeta} 
\sum_{a=1}^b \, \sum_{i=1}^n N_i^a \zeta_i > 0  \, .
\ee
Since \eqref{thebigformula} is continuous at $\gamma_{ij}=0$, the result when some of the 
$\gamma_{ij}$'s vanish can be obtained as a limit of \eqref{thebigformula}.
Furthermore, 
throughout this subsection we shall work with generic FI parameters for which
the left hand side of \refb{econdombeta} never vanishes. In that case 
\refb{thebigformula} is also invariant under small deformations of the FI parameters.

On the Coulomb branch side,  
the Poincar\'e-Laurent polynomial $\QC(\gamma;\zeta;y)$
coincides with its rational counterpart $\bQC
(\gamma;\zeta;y)$. Since
the quiver has no oriented loop there are no contributions
from scaling solutions, therefore we can set
$\Omega_{\rm tot}(\alpha,y)=\OmS(\alpha,y)$ in \refb{essp1spec}. 
We can further set 
$\OmS(\alpha;y)$  
to 0 except when $\alpha$ is equal to one of
the basis vectors $\gamma_\ell$ 
in which case its value is 1. The Coulomb branch formula
\refb{essp1spec} thus reduces to
\be
\label{CouAb}
\QC (\gamma_1+\cdots +\gamma_K; \zeta ;y) = 
\gref(\{\gamma_1,\cdots \gamma_K\};
\{\zeta_1,\cdots \zeta_K\};y)\, .
\ee
It follows from the discussion in \S\ref{sec_noloopany} that the result when some of the 
$\gamma_{ij}$'s vanish can be obtained as a limit of \eqref{CouAb} for generic
non-vanishing $\gamma_{ij}$'s.
Comparing \refb{CouAb} and \refb{thebigformula} we see that the proof of equivalence
of $\QC$ and $\QR$ for Abelian quivers reduces to the proof of equivalence of
$\gref$ and $\gRa$. Furthermore since for both $\gref$ and $\gRa$ the result when some
of the $\gamma_{ij}$'s vanish can be obtained as limits of the result for generic
$\gamma_{ij}$'s, it is enough to prove the equivalence for generic 
non-vanishing $\gamma_{ij}$'s.

This equivalence can now be proved by following exactly the same analysis as
\S 3.3 of ref.\cite{1011.1258}. We summarize the main points, refering
the reader to \cite{1011.1258} for more details.
First, to each permutation $\sigma$, we associate a family of ordered partitions of
$\gamma_1+\cdots \gamma_K$ into vectors $\{\beta^{(a)}\}$ as follows:
\begin{itemize}
\item
Break the sequence $\{\sigma(1), \sigma(2),\cdots \sigma(n)\}$ 
into subsequences $\{\sigma(1), \sigma(2),\cdots \sigma(i_1)\}$,
$\{\sigma(i_1+1),\cdots \sigma(i_2)\}$, $\cdots$, $\{\sigma(i_{a-1}+1),\dots,
\sigma(i_a)\}$, $\cdots$, $0=i_0<i_1<\dots<i_s=K$, such that each subsequence
represents an increasing subsequence: $\sigma(i_{a-1}+1)< \sigma(i_{a-1}+2) <
\cdots < \sigma(i_a)$ for each $a$. $s$ gives
the number of such  increasing subsequences. 
\item  For each such breakup of $\{\sigma(1),\cdots \sigma(K)\}$ into increasing
subsequences, we can associate a partition of $\gamma_1+\cdots \gamma_K$ into
vectors $\beta^{(a)}$ as follows:
\be \label{ehiggscoulomb}
\beta^{(s+1-a)} =  
\sum_{i=i_{a-1}+1}^{i=i_a} \gamma_{\sigma(i)}\ .
\ee
This generates an ordered partition of $\gamma$ but does not necessarily satisfy
\refb{econdombeta}. It is easy to show that with this choice
the power of $y$ in
$y^{\sum_{i<j} \gamma_{\sigma(i)\sigma(j)}}$ 
in \eqref{eindex} matches the power of $y$ inside the sum in
\eqref{thebigformula}
\end{itemize}

For illustration we can consider the permutation $\sigma(1234)
= (3412)$. The increasing subsequences
are $\{\{34\},$ $\{12\}\}$, $\{\{3\}, \{4\}, \{12\}\}$,
$\{\{34\}, \{1\}, \{2\}\}$, and $\{\{3\},\{4\}, \{1\}, \{2\}\}$.
Associated partitions are
$\{\gamma_1+\gamma_2,\gamma_3+\gamma_4\}$, 
$\{\gamma_1+\gamma_2,\gamma_4, \gamma_3\}$,
$\{\gamma_2,\gamma_1,\gamma_3+\gamma_4\}$
and
$\{\gamma_2,\gamma_1,\gamma_4,\gamma_3\}$, respectively.
For each of these partitions we have 
\be
2\sum_{a\le b} \sum_{j<i}\gamma_{ji} N^a_i N^b_j - \sum_{1\le i<j\le 4}
\gamma_{ij} =\sum_{1\le i<j \le 4} \gamma_{\sigma(i)\sigma(j)}
= \gamma_{12}+\gamma_{34}-\gamma_{13}-\gamma_{14}-\gamma_{23}-\gamma_{24}
\, .
\ee

The above analysis shows how to each permutation $\sigma$ we can associate
a set of ordered partitions of $\gamma_1,\cdots \gamma_K$ into vectors 
$\beta^{(1)},\cdots \beta^{(s)}$. The converse is also true -- 
all ordered partitions of the vectors $(\gamma_1,\dots,
\gamma_K)$,  before imposing \eqref{econdombeta},  
are in  one to one correspondence with the set of all 
increasing subsequences of all the
permutations of $(12\dots n)$  via 
the rule \eqref{ehiggscoulomb}.
For a given permutation we shall call an 
increasing subsequence maximal
if it is not possible to build bigger 
 increasing
subsequences involving the elements of the subsequence.
In the example described above the maximal increasing subsequences are
$\{\{34\}, \{12\}\}$.
The complete
contribution to the coefficient of $y^{\sum_{i<j} \gamma_{\sigma(i)\sigma(j)}}$
in the Reineke formula can be generated 
by beginning with the maximal increasing subsequences
associated with the permutation $\sigma$ and combining them
with the contribution from other increasing subsequences
associated with the same permutation,
but we must be careful to pick only those partitions which
satisfy \refb{econdombeta}. As shown in \cite{1011.1258},
a given permutation contributes if and only if its
maximal increasing subsequences generate a
partition via
\eqref{ehiggscoulomb} satisfying \eqref{econdombeta}, and none of the other
(non-maximal) increasing subsequences generate
an allowed partition.\footnote{In particular, if some non-maximal increasing 
subsequence satisfies \eqref{econdombeta}, then the net contribution from
the partitions associated with the maximal and non-maximal increasing subsequences
of a given permutation cancel pairwise.}
The sign of the contribution is
given by $(-1)^{s-1}$ where $s$ is the number of maximal increasing
subsequences.

It is now easy to see that the condition that the maximal increasing subsequences
generate a partition satisfying \eqref{econdombeta} translates to
\be
\sum_{i=k+1}^K \dx_{\sigma(i)} > 0 \quad \hbox{for} \quad \gamma_{\sigma(k),\sigma(k+1)} < 0\, ,
\ee
and the condition that none of the non-maximal increasing subsequences satisfy
\refb{econdombeta} translates to
\be
\sum_{i=k+1}^K \dx_{\sigma(i)} < 0 \quad \hbox{for} \quad \gamma_{\sigma(k),\sigma(k+1)} > 0\, .
\ee
These precisely agree with \refb{esign}. Finally the number of maximal increasing
subsequences is one more than the number of negative $\gamma_{\sigma(k),\sigma(k+1)}$. 
Thus we
have
\be 
(-1)^{s-1} = \prod_{k=1}^{K-1} \, {\rm sign}(\gamma_{\sigma(k),\sigma(k+1)})\, ,
\ee
in agreement with \refb{esign2}. This proves that the Reinecke formula
\eqref{thebigformula} agrees with the Coulomb index \eqref{eindex1},
and hence also the Coulomb branch formula \refb{CouAb} for the
Poincar\'e-Laurent polynomial\footnote{This relation
holds away from walls of marginal and threshold stability. 
On such a wall, the Higgs branch index on the l.h.s.
given by \refb{thebigformula}, \refb{econdombeta}, 
is still well defined but the derivation of the Coulomb index given in \refb{esign},
\refb{esign2}
clearly breaks down. Furthermore even the Higgs branch index
given by \refb{thebigformula}, \refb{econdombeta}
does not have 
the form of a
symmetric Laurent polynomial, and hence cannot be interpreted as ${\rm Tr}'(-y)^{2J_3}$
associated with a quantum system. On a threshold stability wall we shall define
$\gref$ by deforming the FI parameters away from the wall since there is no
independent definition of $\gref$ in this case, and the result is independent of the
deformation. However for $\gRa$ we shall continue to use the definition given in
\refb{thebigformula}, \refb{econdombeta} for reasons which will become clear in
\S\ref{snonprimitive}. The price we pay is that $\gref$ and $\gRa$ are no longer
equal on the thereshold stability wall.
\label{fo1}},
\ben \label{equality}
\gRa(\{\gamma_1,\cdots \gamma_K\}; \{\zeta_1, \cdots \zeta_K\};y)
&=& \gref(\{\gamma_1,\cdots \gamma_K\}; \{\zeta_1,\cdots \zeta_K\};y) \, ,
\nonumber \\
\QR(\gamma_1+\cdots +\gamma_K; \zeta ;y)
&=& \QC(\gamma_1+\cdots + \gamma_K; \zeta;y)\, .
\nonumber \\
\een
For charge configurations relevant for wall-crossing, where all the $\gamma_\ell$'s lie
in a plane and the $\zeta_\ell$'s are determined in terms of $\gamma_{ij}$, 
this equivalence was proved by 
induction in \cite{1112.2515}.
Here we have shown that it holds for any Abelian quiver without
oriented  loop.

\subsection{Non-Abelian quivers with primitive dimension
vector and without loops}  \label{sec_equivnonAbelian}

For non-Abelian quivers without loops we still can set 
$\Omega_{\rm tot}(\alpha;y)=\OmS(\alpha;y)$.
We also have
$\OmS(\alpha;y)=0$ unless
$\alpha=\gamma_\ell$, and
$\OmS(\gamma_\ell;y)=1$.
Thus $\OmS (\alpha_i/m_i;y^{m_i})$ is non-vanishing 
(and equal to 1) only when
$\alpha_i/m_i$ is a basis vector. Furthermore
for primitive charge vector, $\QC(\gamma; \zeta;y)$ is equal to
$\bQC(\gamma; \zeta;y)$.
In that case we can express \refb{essp1} as
\ben \label{ecoulombexp}
\QC\left(\sum_i N_i \gamma_i; \zeta;y\right) =\sum_{\{k^{(\ell)}_j\}\atop \sum_\ell \ell k^{(\ell)}_j=N_j}
&& \gref( \{ (\ell\gamma_j)^{k^{(\ell)}_j}\}; \{ (\ell \zeta_j)^{k^{(\ell)}_j}\}; y)
\nonumber \\ &&
\times \left\{ \prod_{j=1}^K 
\prod_\ell {1\over k^{(\ell)}_j!}\left( {y - y^{-1}\over \ell (y^\ell - y^{-\ell})}\right)^{k^{(\ell)}_j}
\right\}\, .
\een
The argument of $\gref$ given above corresponds to a choice
of $\{\alpha_1,\cdots \alpha_n\}$ where we have $k^{(\ell)}_j$ copies of the vector
$\ell\gamma_j$ for $\ell=1,2,3,\cdots$, $j=1,\cdots K$,
and the FI parameter associated with each such copy 
is
$\ell \zeta_j$. 
It is understood that if some of the $\alpha_{ij}$'s appearing in the argument of
$\gref$ vanish then we must deform them in a way so that the deformed quiver
does not contain any oriented loop, compute $\gref$ and then take the deformations
back to zero.
The $\prod_i \prod_\ell {k^{(\ell)}_i!}$ factor in the denominator
represents $|{\rm Aut}(\{\alpha_1, \alpha_2,\cdots, \alpha_n\})|$. 

On the other hand, it was shown in \cite{1011.1258,ReinekeStoppaWeist} 
that the stack invariants \eqref{ec5}
satisfy the Abelianization formula\footnote{In \cite{1011.1258}, appendix D we proved
this formula for a special choice of  FI parameters
relevant for wall crossing. However this  assumption was inessential and the result
easily generalizes to the case of general FI parameters\cite{ReinekeStoppaWeist}.
The formula \eqref{MPSform} also holds in the case of quivers with loops, but we shall
not make use of this fact here.
}
\be
\label{MPSform}
\gR(\{N_j\}; \{\gamma_j\};\{\zeta_j\};y) = \sum_{\{k_\ell\}\atop
\sum_\ell \ell\, k_{\ell}=N_i } 
\gR(\{\hat N_j\}; \{\hat\gamma_j\};\{\hat\zeta_j\};y) 
\prod_{\ell} \left[ \frac{1}{k_\ell!} \left( \frac{ (y-1/y) }{\ell(y^\ell-y^{-\ell})}\right)^{k_\ell} \right] 
\ee
where, on the r.h.s, $\gR(\{\hat N_j\}; \{\hat\gamma_j\};\{\hat\zeta_j\};y)$ 
is the stack invariant of a quiver $\hat\cQ$ defined as follows: 
\begin{itemize}
\item The vertices of $\hat\cQ$ are obtained by replacing the vertex $i$ of $\cQ$ by a
collection $i_{\ell,k}$ of vertices with $k=1,\dots k_\ell$ for any $\ell$ in the partition 
$N_i=\sum_\ell \ell\, k_{\ell}$, and keeping all the other vertices $j\neq i$ of $\cQ$;
\item Each arrow $j\to k$ in $\cQ$ with $j,k\neq i$ induces an arrow $j\to k$ in $\hat\cQ$;
\item The arrows $i\to j$ in $\cQ$ (resp. $j\to i$) with $j\neq i$ induce $\ell$ arrows $i_{\ell,k}\to j$ (resp. $j\to i_{\ell,k}$) in $\hat\cQ$
for each $\ell,k$;
\item {\it The dimension at each node $i_{\ell,k}$ in $\hat\cQ$ is equal to one, hence justifying the name}; the dimensions
at the other nodes $j\neq i$ are the same as the dimensions $N_j$ in the original quiver $\cQ$;
\item The FI parameter at each node $i_{\ell,k}$ in $\hat\cQ$ is equal to $\ell \zeta_i$;
the FI parameters at the other nodes $j\neq i$ are the same as the FI parameters $\zeta_j$ in the original quiver $\cQ$.
\end{itemize}
Thus for example on the right hand side of \refb{MPSform} the set
$\{\hat N_j\}$ will include all the $N_j$'s except $N_i$ and $\sum_\ell k_\ell$
number of 1's, the  set $\{\hat \gamma_j\}$ will include all the $\gamma_j$'s 
except $\gamma_i$, $k_1$ copy of $\gamma_i$, $k_2$ copy of $2\gamma_i$,
etc., and the set $\{\hat\zeta_i\}$ will include all the $\zeta_j$'s except $\zeta_i$,
$k_1$ copies of $\zeta_i$, $k_2$ copies of $2\zeta_i$ etc. 
By successive application of this formula we can express the stack invariant
of any non-Abelian quiver in terms of that of a family of Abelian quivers: 
\be \label{ehiggsexp}
\gR(\{N_i\}; \{\gamma_i\};\{\zeta_i\};y) =\!\!\!\!\!\!\!\!
\sum_{\{k^{(\ell)}_j\}\atop \sum_\ell \ell k^{(\ell)}_j=N_j}\!\!\!\!\!\!\!\!
\gRa( \{ (\ell\gamma_j)^{k^{(\ell)}_j}\}; \{ (\ell \zeta_j)^{k^{(\ell)}_j}\}; y)
\prod_{i=1}^K 
\prod_\ell {1\over k^{(\ell)}_i!}\left( {y - y^{-1}\over \ell (y^\ell - y^{-\ell})}\right)^{k^{(\ell)}_i}\, .
\ee
Using \refb{ec5a} we can replace the left hand side by
$\QR\left(\sum_{i=1}^K N_i \gamma_i;\zeta;y\right)$. The resulting equation
is identical to 
\refb{ecoulombexp} with $\QC$
replaced by $\QR$ and
$\gref$ replaced by $\gRa$.
Using the equivalence of $\gref$ and $\gRa$ proven in \refb{equality},
we get
\be 
\QR \left(\sum_\ell N_\ell \gamma_\ell; \zeta; y\right) 
= \QC\left(\sum_\ell N_\ell \gamma_\ell; \zeta; y\right)\,.
\ee
Thus
the Higgs and Coulomb branch formulae are equivalent for non-Abelian quivers as well,
so long as the dimension vector $\{N_i\}$ is primitive.

\subsection{Non-primitive dimension vector} \label{snonprimitive}

We now turn to the case of a non-Abelian quiver with 
non-primitive dimension vector.
In this case the Coulomb branch formula is still given by
\refb{essp1} although $\QC$ and $\bQC$ are no longer identical.
On the other hand the Higgs branch formula is given by 
\refb{eq:inversestackinv} (or equivalently
 \refb{gtoQb}) and
\refb{ec5}. Our goal in this section will be to prove the equality of
$\QC$ and $\QR$.

When the dimension vector $\{N_i\}_{i=1\dots K}\equiv \vec N$ 
is not primitive, the Abelianization formula
\eqref{ehiggsexp} is still known to hold, and can be used to express
$\gR$ in terms of $\gRa\big(\{(\ell \gamma_j)^{k^{(\ell)}_j}\}; $ $\{(\ell \zeta_j)^{k^{(\ell)}_j}\};y\big)$
for integers $k^{(\ell)}_j$ satisfying
$\sum_{\ell} \ell k^{(\ell)}_j = N_j$.
However some of the Abelian stack invariants 
$\gRa(\{\hat\gamma_i\} ;\{\hat\zeta_i\};y)$ appearing on the r.h.s. have stability conditions 
$\{\hat\zeta\}$ lying on walls of threshold stability.
This happens when the set $\{\hat \gamma_i\}$ can be divided into
several subsets $A_1,A_2,\cdots$ such that $\sum_{i\in A_s}\hat\gamma_i\parallel
\sum_\ell N_\ell\gamma_\ell$ for each $s$.
As a result $\gRa(\{\hat\gamma_i\} ;\{\hat\zeta_i\};y)$ can no longer be
equated to $\gref(\{\hat\gamma_i\} ;\{\hat\zeta_i\};y)$ (see footnote \ref{fo1}). 
Let $\{\hat\zeta_{i}\}'$ be a sufficiently generic perturbation of the FI parameters away 
from the wall and sufficiently small so that
no other walls of marginal stability are crossed.
In this case $\gRa(\{\hat\gamma_i\} ;\{\hat\zeta_i\}';y)$ can again be
equated to $\gref(\{\hat\gamma_i\} ;\{\hat\zeta_i\}';y)$ and furthermore will be 
independent of the deformation. Now the HN recursion method \cite{Harder:1975, MR1974891} relates  $\gRa(\{\hat\gamma_i\} ;\{\hat\zeta_i\}';y)$ and 
$\gRa(\{\hat\gamma_i\} ;\{\hat\zeta_i\};y)$ as follows:
\be
\label{eq:zetatozetap}
\begin{split}
\gRa&(\alpha_1, \cdots \alpha_n; c_1,\cdots c_n; y) \\
=&\sum_{k\geq 1} (y^{-1}-y)^{1-k}\,  \!\!\!\!\!\!\!\!\!\!\!\!\!\!\!\!
\sum_{\substack{\{A_s\}\\ 
\sum_{i\in A_s} c_i=0 \, \, \forall \, s; \, \,
\cup_{s=1}^k A_s=\{1,2,\cdots n\}}}  \!\!\!\!\!\!\!\!
\prod_{s=1}^k \gRa(\{\alpha_i; \, \, i\in A_s\}; \{c_i'; \,\, i\in A_s\}; y) 
\end{split}
\ee
where the sum over $\{A_s\}$ runs over all {\it unordered partitions} of the integers
 $1,2,\cdots n$ into sets $A_1, A_2,\cdots A_k$ subject to the conditions
 indicated above.  The symbol $\{c_i'\}$ denotes a generic 
deformation in which the FI parameters
associated with each of the $n$ nodes are deformed independently so that the configuration
moves away from the wall of threshold stability,
subject to the constraint that the sum of the FI parameters carried by all the nodes
must vanish.  
It was shown in Ref. \cite{Moz:2012}  that the combinatorics of the
summation in (\ref{eq:zetatozetap}) can be summarized
by an equality of generating functions, which in our notation reads
\be
\label{MRformula}
\begin{split}
 1& + (y^{-1}-y)^{-1} \sum_{\{k^{(\ell)}_{i}\}, \sum_{\ell} \ell k^{(\ell)}_{i}\propto N_i}
\gRa\left(\{(\ell \gamma_j)^{k^{(\ell)}_j}\}; \{(\ell \zeta_j)^{k^{(\ell)}_j}\};y\right)
\prod_{i,\ell}{ (t_{i,\ell})^{k^{(\ell)}_{i}}\over  k^{(\ell)}_{i}!}  \\
=& \exp\Big[
(y^{-1}-y)^{-1} \sum_{\{k^{(\ell)}_{i}\}, \sum_{\ell} \ell k^{(\ell)}_{i}\propto N_i}
\gRa\left(\{(\ell \gamma_j)^{k^{(\ell)}_j}\}; \{(\ell \zeta_j)^{k^{(\ell)}_j}\}';y\right)
\prod_{i,\ell}{ (t_{i,\ell})^{k^{(\ell)}_{i}}\over  k^{(\ell)}_{i}!}
\Big] \, ,
\end{split}
\ee
where $t_{i,\ell}$ are formal parameters and the sum over $\{k_i^{(\ell)}\}$ on either
side runs over all integers $k_i^{(\ell)}$ for which $\sum_\ell k_i^{(\ell)}$ is
of the form $cN_i$ for some fixed vector $N_i$ and arbitrary constant $c$. 
Note that for the deformed stability conditions $\zeta'$,
$\gRa$ can be equated to $\gref$, which  is  by definition equal to the value of $\gref$ for
the undeformed stability condition $\zeta$ (see footnote \ref{fo1}). As a result, we get for $\gRa$ either from (\ref{eq:zetatozetap}) directly, or after
equating coefficients of  $\prod_{i,\ell} (t_{i,\ell})^{k^{(\ell)}_{i}}$ on either side,
\be
\begin{split}
\label{esoln}
\gRa&\left(\{(\ell \gamma_j)^{k^{(\ell)}_j}\}; \{(\ell \zeta_j)^{k^{(\ell)}_j}\};y\right)=
 \Big(\prod_{i,\ell}{ k^{(i)}_{\ell}!} \Big)  \sum_k  {(y^{-1}-y)^{1-k}  \over k!} 
\\
&\times  \!\!\!\!
\sum_{\{k^{(\ell,s)}_i,\, s=1,\cdots k\} \atop \sum_{\ell} \ell k^{(\ell,s)}_i\propto 
\sum_\ell \ell\, k^{(\ell)}_i
 \, \forall \, s, \,
\sum_{s}  k^{(\ell,s)}_i =  k^{(\ell)}_i} \!\!\!\!\!
\prod_s \bigg\{
\gref\left(\{(\ell \gamma_j)^{k^{(\ell,s)}_j}\}; \{(\ell \zeta_j)^{k^{(\ell,s)}_j}\};y\right)
\prod_{i,\ell}{ 1\over  k^{(\ell,s)}_i!} \bigg\}\, ,
\end{split}
\ee
where the sum over $\{k^{(\ell,s)}_i\}$ runs over all {\it ordered} partitions of
$\{k^{(\ell)}_i\}$. 
Using this we can express the Abelianization formula \refb{ehiggsexp} as
\be
\label{soln2}
\begin{split}
\gR&(\{N_i\};\{\gamma_i\}; \{\zeta_i\}; y) \\
=&\sum_{\{k^{(\ell)}_j\}\atop \sum_\ell \ell k^{(\ell)}_j=N_j}
\Big(\prod_{i,\ell}{ k^{(i)}_{\ell}!}\Big) 
\sum_k  { (y^{-1}-y)^{1-k} \over k!}
 \sum_{\{k^{(\ell,s)}_i, 1\le s\le k\}\atop \sum_{\ell} \ell k^{(\ell,s)}_i\propto \sum_\ell \ell\, k^{(\ell)}_i
 \, \forall \, s, \,
\sum_{s} k^{(\ell,s)}_i = k^{(\ell)}_i}
\prod_{s=1}^k \bigg\{
\\ & 
\gref\left(\{(\ell \gamma_j)^{k^{(\ell,s)}_j}\}; \{(\ell \zeta_j)^{k^{(\ell,s)}_j}\};y\right)
\prod_{i,\ell}{ 1\over  k^{(\ell,s)}_i!} \bigg\}
\times \left\{ \prod_{i=1}^K 
\prod_\ell {1\over k^{(\ell)}_i!}\left( {y - y^{-1}\over \ell (y^\ell - y^{-\ell})}\right)^{k^{(\ell)}_i}
\right\}\, . 
\end{split}
\ee
We can now remove the sum over $\{k^{(\ell)}_j\}$ by relaxing the constraint 
$\sum_{s}  k^{(\ell,s)}_i =  k^{(\ell)}_i$, replacing $k^{(\ell)}_i$ by
$\sum_s k^{(\ell,s)}_i$ everywhere, and imposing the constraints
$\sum_{\ell, s} \ell k^{(\ell,s)}_i=N_i$ and 
$\sum_{\ell} \ell k^{(\ell,s)}_i \propto N_i$. This gives
\be
\label{soln2a}
\begin{split}
\gR&(\{N_i\};\{\gamma_i\}; \{\zeta_i\}; y)  \\
=&\sum_k  {(y^{-1}-y)^{1-k}  \over k!} \!\!\!\!\!
 \sum_{\{k^{(\ell,s)}_i, 1\le s\le k\}\atop \sum_{\ell} \ell k^{(\ell,s)}_i \propto N_i
 \, \forall \, s, \,
\sum_{s,\ell} \ell \, k^{(\ell,s)}_i = N_i}
\prod_{s=1}^k \Bigg[
\gref\left(\{(\ell \gamma_j)^{k^{(\ell,s)}_j}\}; \{(\ell \zeta_j)^{k^{(\ell,s)}_j}\};y\right)
\\ &
\times \left\{ \prod_{i=1}^K 
\prod_\ell {1\over  k^{(\ell,s)}_i!}\left( {y - y^{-1}\over \ell (y^\ell - y^{-\ell})}\right)^{k^{(\ell,s)}_i}
\right\} \Bigg]\, . 
\end{split}
\ee

Following the discussion at the beginning of \S\ref{sec_equivnonAbelian} but without imposing the
primitivity condition on the dimension vector $\gamma\equiv \sum_\ell
N_\ell \gamma_\ell$, we see that 
the Coulomb branch formula \refb{essp1} takes the form
\be \label{erenon}
\begin{split}
\QC(\gamma; \zeta;y) =& \sum_{m|\gamma} 
\frac{\mu(m)}{ m}  {y - y^{-1}\over y^m - y^{-m}}
\bQC(\gamma/m; \zeta;y^m) \\
\bQC(\gamma; \zeta;y) =&
\sum_{\{k^{(\ell)}_j\}\atop \sum_\ell \ell k^{(\ell)}_j=N_j}\!\!\!
\gref( \{ (\ell\gamma_j)^{k^{(\ell)}_j}\}; \{ (\ell \zeta_j)^{k^{(\ell)}_j}\}; y)
 \prod_{j=1}^K 
\prod_\ell {1\over k^{(\ell)}_j!}\left( {y - y^{-1}\over \ell (y^\ell - y^{-\ell})}\right)^{k^{(\ell)}_j}\ .
\end{split}
\ee
Let us now define $\gC$ by a formula
analogous to \refb{gtoQb} with $\bQR$ replaced by $\bQC$ on the right hand side:
\be
\label{gtoQbcoulomb}
\begin{split}
 \gC&(\{N_1,\cdots N_K\}; \{\gamma_1,\cdots \gamma_K\};
\{\zeta_1,\cdots \zeta_K\};y)  \\
=& \sum_k \sum_{\substack{\{\alpha_i\}\\ \sum_{i=1}^k \alpha_i=\sum_\ell N_\ell \gamma_\ell\\
\alpha_i\parallel \sum_{N_\ell \gamma_\ell}\,\mathrm{for}\,\, i=1,\dots, k} }
\, \frac{(-1)^{k-1}}{k! (y-y^{-1})^{k-1}}
\prod_{i=1}^k\, \bQC(\alpha_i;\zeta;y)\, .
\end{split}
\ee
Proving the equivalence of $\QR$ and $\QC$ is then equivalent to proving the
equivalence of $\gR$ and $\gC$. 
Now using \refb{erenon} we get
\be \label{esoln3}
\begin{split}
 \gC&(\{N_1,\cdots N_K\}; \{\gamma_1,\cdots \gamma_K\};
\{\zeta_1,\cdots \zeta_K\};y) \\
=& \sum_k \sum_{\substack{\{\alpha_s\} \\
\sum_{s=1}^k \alpha_s=\sum_i N_i \gamma_i\\
\alpha_s\parallel \sum_{N_i \gamma_i}\,\mathrm{for}\,\, s=1,\dots, k}}
\, \frac{(-1)^{k-1}}{k! (y-y^{-1})^{k-1}} \prod_{s=1}^k  
\sum_{\{k^{(\ell,s)}_j\}\atop \sum_\ell \ell k^{(\ell,s)}_j\gamma_j=\alpha_s}\!\!\!\!\!\!
\gref( \{ (\ell\gamma_j)^{k^{(\ell,s)}_j}\}; \{ (\ell \zeta_j)^{k^{(\ell,s)}_j}\}; y)
 \\ &
\times \left\{ \prod_{i=1}^K 
\prod_\ell {1\over k^{(\ell,s)}_i!}\left( {y - y^{-1}\over \ell (y^\ell - y^{-\ell})}\right)^{k^{(\ell,s)}_i}
\right\}\, ,
\end{split}
\ee
where the sum over $\{k^{(\ell,s)}_i\}$ runs over all {\it ordered} partitions of
$\{k^{(\ell)}_i\}$. 
We can now remove the sum over $\alpha_s$'s by relaxing the constraint that
$\sum_\ell \ell k^{(\ell,s)}_j\gamma_j=\alpha_s$, but imposing the
constraints $\sum_{\ell, s} \ell k^{(\ell,s)}_i=N_i$ and 
$\sum_{\ell} \ell k^{(\ell,s)}_i \propto N_i$.  This gives
\be \label{esoln3a}
\begin{split}
 \gC&(\{N_1,\cdots N_K\}; \{\gamma_1,\cdots \gamma_K\};
\{\zeta_1,\cdots \zeta_K\};y)  \\
=& \sum_k  {1\over k!} \, (y^{-1}-y)^{1-k} \!\!\!\!\!
 \sum_{\{k^{(\ell,s)}_i, 1\le s\le k\}\atop \sum_{\ell} \ell k^{(\ell,s)}_i \propto N_i
 \, \forall \, s, \,
\sum_{s,\ell} \ell \, k^{(\ell,s)}_i = N_i}
\prod_{s=1}^k \Bigg[
\gref\left(\{(\ell \gamma_j)^{k^{(\ell,s)}_j}\}; \{(\ell \zeta_j)^{k^{(\ell,s)}_j}\};y\right)
\\ &
\times \left\{ \prod_{i=1}^K 
\prod_\ell {1\over  k^{(\ell,s)}_i!}\left( {y - y^{-1}\over \ell (y^\ell - y^{-\ell})}\right)^{k^{(\ell,s)}_i}
\right\} \Bigg]\, . 
\end{split}
\ee
Since the right hand side matches the right hand side of \refb{soln2a}, it follows that 
$\gR$ and $\gC$ are identical, and therefore the Higgs and Coulomb branch computations
$\QR$ and $\QC$ are equivalent, even for non-primitive dimension vector. 

Given the
equivalence of $\QR$ (defined by the Harder-Narasimhan
recursion \refb{eq:inversestackinv}, \refb{ec5}) and 
 $\QC$ (defined by the Abelianization formula \refb{erenon}),
 and given the equivalence of the Abelian indices
 $\gref$ and $\gRa$ for deformed FI parameters,  
we conclude that the Poincar\'e-Laurent polynomial $\QR$ of a quiver without
loop  satisfies, for arbitrary dimension vector, 
the Abelianization formula: 
\be\label{erenonhiggs}
\begin{split}
\QR(\gamma; \zeta;y) =& \sum_{m|\gamma} 
\frac{\mu(m)}{ m}  {y - y^{-1}\over y^m - y^{-m}}
\bar Q_{\rm Higgs}(\gamma/m; \zeta;y^m)   \\
\bar Q_{\rm Higgs}(\gamma; \zeta;y) =&
\sum_{\{k^{(\ell)}_j\}\atop \sum_\ell \ell k^{(\ell)}_j=N_j}\!\!\!\!
\gRa( \{ (\ell\gamma_j)^{k^{(\ell)}_j}\}; \{ (\ell \zeta_j)^{k^{(\ell)}_j}\}'; y)
\,\prod_{j=1}^K 
\prod_\ell {1\over k^{(\ell)}_j!}\left( {y - y^{-1}\over \ell (y^\ell - y^{-\ell})}\right)^{k^{(\ell)}_j}
\ .
\end{split}
\ee
This can be regarded as the main result of this subsection.

\section*{Acknowledgments}
We are grateful to Frederik Denef and Markus Reineke  for valuable discussions.
The work of A.S.  was
supported in part by the project 11-R\&D-HRI-5.02-0304
and the J. C. Bose fellowship of 
the Department of Science and Technology, India. 

\section*{Note added in arxiv:v2 (after publication)}

It should be clear from  \S 2 that there are many 
possible deformations of the $\ta_{ij}$'s, leading to different recursion relations.  
The final result for $F$ will of course be the same, but some may be more efficient than
others. Here we  present an alternative recursion, which appears to be 
more efficient than \eqref{efinal1}.
For this we scale $\ta_n$ by $\lambda$ and take
$\lambda$ to 0 keeping the $\cc_k$'s fixed. In the limit $\lambda\to 0$
the $n$-th center can be treated as a probe moving in the background
of the other centers and the result is
\be \label{esum1}
\Theta\left(-\ta_{n-1, n} \cc_n\right) (-1)^{\Theta(-\ta_{n-1,n})}
F(\{\ta_1,\cdots \ta_{n-1}\};  \{\cc_1, \cdots \cc_{n-2}, \cc_{n-1}+\cc_n\})\, .
\ee
During this deformation we also pick up contribution from the collinear scaling
solutions. Since only $\ta_n$ is deformed, any collinear scaling solution
that appears during the deformation  must include the $n$-th center and
hence the only possible configurations are those involving the centers 
$k+1, k+2, \cdots n$ for $0\le k\le n-3$. They occur at
\be
\lambda_k = - \sum_{i,j\atop k+1\le i<j\le n-1} \ta_{ij} / \sum_{i=k+1}^{n-1} \ta_{in}
\ee
provided
\be 
\sum_{i,j\atop k+1\le i<j\le n-1} \ta_{ij} \sum_{i,j\atop k+1\le i<j\le n} \ta_{ij} <0
\ee
so that $\lambda_k$ lies between 0 and 1. The net contribution from these
scaling solutions can be computed as before, leading to 
\be
 \label{enewrecur}
\begin{split}
F&(\{\ta_1,\cdots \ta_n\};  \{\cc_1, \cdots \cc_n\}) 
=\Theta\left(- \ta_{n-1, n} \cc_n\right) (-1)^{\Theta(-\ta_{n-1,n})}
F(\{\ta_1,\cdots \ta_{n-1}\};  \{\cc_1, \cdots \cc_{n-1}\}) \\
& + \sum_{k=2}^{n-3} F(\{\ta_1, \cdots \ta_k, \ta_{k+1}+\cdots \ta_{n-1}+\lambda_k \ta_n\};
\{\cc_1, \cdots \cc_k, \cc_{k+1}+\cdots \cc_n\}) \qquad \qquad \qquad \\
& \times G(\ta_{k+1}, \cdots \ta_{n-1}, \lambda_k \ta_n) \, \Theta\left(-
\sum_{i,j\atop k+1\le i<j\le n-1} \ta_{ij} \sum_{i,j\atop k+1\le i<j\le n} \ta_{ij}\right)
{\rm sign}\left(\sum_{i=k+1}^{n-1} \ta_{kn}\right) \, . 
\end{split}
\ee
This recursion is implemented in the {\tt CoulombHiggs.m} package (v2.0) and used
by default. The old recursion  \eqref{efinal1} can be used by  
setting  \var{{\tt \$QuiverRecursion}} to 0.

\appendix

\section{The {\sc Mathematica} package ``{\tt CoulombHiggs}"}

To facilitate further investigation, we provide  a {\sc Mathematica} package allowing to compute the Coulomb index $\gref$ for multi-centered black holes and  the Poincar\'e-Laurent 
polynomial $\QC(\gamma;\zeta;y)$ and $\QR(\gamma;\zeta;y)$ 
of quiver moduli spaces using the Coulomb branch and
Higgs branch formulae.
We also provide three example files where this package
is used to evaluate quiver invariants for the Kronecker quiver ({\tt Kronecker.nb}),
for non-Abelian 3-node quivers ({\tt Threenode.nb}) and for several 4 and 5-node Abelian
quivers considered in \cite{Manschot:2012rx} ({\tt Multinode.nb}). The validity of the algorithm
for the $F$ and $G$ indices is tested in a fourth file {\tt CoulombIndexCheck.nb}. 
All these files are included in
the ``source'' of this paper available from arXiv and can be obtained from the second name authors' webpage.

Assuming that the file {\tt CoulombHiggs.m} is present in the user's {\sc Mathematica} Application 
directory, the package is loaded by entering 

\mathematica{0.9}{ <<CoulombHiggs`}{CoulombHiggs v 2.0 - A package for evaluating quiver invariants using the Coulomb and Higgs branch formulae. }

If the file  {\tt CoulombHiggs.m} has not yet been copied in the user's {\sc Mathematica} Application 
directory but is in the same directory as the notebook, evaluate instead

\mathematica{0.9}{SetDirectory[NotebookDirectory[]]; <<CoulombHiggs`}{CoulombHiggs v2.0 - A package for evaluating quiver invariants using the Coulomb and Higgs branch formulae.}

The first main routine  is {\tt \color{functioncolor} CoulombBranchFormula}, whose basic usage is illustrated below: \footnote{Note the following changes in v2.0: the fugacity $y$ 
is no longer a parameter of \fun{CoulombBranchFormula} and \fun{QuiverBranchFormula},
and the former computes the Dolbeault polynomial in terms of $\OmS(\alpha_i,t)$, 
rather than expressing  the Poincar\'e polynomial in terms of $\OmS(\alpha_i)$.
Other changes are highlighted by margin notes below. }

\mathematica{0.9}{Simplify[CoulombBranchFormula[4\{\{0, 1, -1\},\{-1, 0, 
   1\}, \{1, -1, 0\}\}, \{1/2, 1/6, -2/3\}, \{1, 1, 1\}]]  
     }
   {$ 2+ \frac{1}{y^2}+y^2 +  \text{OmS}(\{1,1,1\},y,t) $
}


This routine computes the Dolbeault polynomial \eqref{edolbeault} of the quiver moduli space, expressed in terms of the single-centered indices. 
The first argument corresponds to the matrix of DSZ products $\alpha_{ij}$ (an antisymmetric matrix of integers), the second to the FI parameters $\zeta_i$ (a vector of rational numbers), the third to the dimension vector $N_i$ (a vector of integers). 
The variables
 $y$ and $t$ are fugacities conjugate to the sum of the Dolbeault
degrees $p+q$ (i.e. the angular
momentum) and to the difference of the Dolbeault
degrees $p-q$, respectively. The Poincar\'e-Laurent
polynomial is obtained by setting $t=1$. For generic superpotential, the single-centered
indices $\OmS(\gamma,y,t)$ are conjectured to be  
independent of $y$.  In the above example,  the 
Dolbeault polynomial of the moduli space of a three-node Abelian cyclic quiver with $4$ arrows between each subsequent node is expressed in terms of the single-centered index
$\OmS(\gamma_1+\gamma_2+\gamma_3,y,t)$. 
The second main routine is {\tt \color{functioncolor} HiggsBranchFormula}, which computes the
Poincar\'e-Laurent polynomial  using the Higgs branch formula \eqref{eq:inversestackinv} (which is only valid for quivers without oriented  loop, but the routine works irrespective of this
assumption). The arguments are the same as for {\tt \color{functioncolor} CoulombBranchFormula}: 

\mathematica{0.9}{Simplify[HiggsBranchFormula[\{\{0, 3\},\{-3, 0\}\}, \{1/2,-1/2\}, \{2, 2\}]]
  }
   {$ -\frac{\left(y^2+1\right) \left(y^8+y^4+1\right)}{y^5} $
}

The above command computes the Poincar\'e-Laurent polynomial for the Kronecker quiver with 3 arrows, FI parameters $(1/2,-1/2)$, dimension vector $(2,2)$. The package allows for much
more, as documented below. 
Inline documentation can be 
obtained by typing e.g. 

\mathematica{0.9}{?CoulombBranchFormula}
   {}

\subsection{Symbols}

\defvar{y}{fugacity conjugate to the sum of Dolbeault degrees $p+q$ (i.e. angular momentum);}

\defvar{t}{fugacity conjugate to the difference of Dolbeault degrees $p-q$;}

\defn{Om}{\vardef{charge vector},\newdefy}{denotes the refined index  
$\Omega(\gamma,y)$;}

\defn{Omb}{\vardef{charge vector},\newdefy }{denotes the rational refined index
$\bar\Omega(\gamma,y)$;}

\defn{OmS}{\vardef{charge vector},\newdefy,\vardef{t} }{denotes the single-centered index 
$\OmS(\gamma,y,t)$. } \newnote 

\defn{OmS}{\vardef{charge vector},\newdefy}{denotes 
$\OmS(\gamma,y)\equiv \OmS(\gamma,y,t=1)$.}

\defn{OmS}{\vardef{charge vector}}{denotes 
$\OmS(\gamma,y)$, under the assumption that it is independent of $y$ (which is
conjectured to be the case for  generic superpotential)}

\defn{OmT}{\vardef{charge vector},\newdefy }{denotes the (unevaluated)  
function $\Omega_{\rm tot}(\gamma,y)$ defined in \refb{essp2};}

\defn{Coulombg}{\vardef{list of charge vectors},\newdefy }{: denotes the (unevaluated) Coulomb index $\gref(\{\alpha_i\},\{c_i\},y)$, leaving the FI parameters unspecified;} 

\defn{HiggsG}{\vardef{charge vector},\newdefy }{denotes the (unevaluated)  
stack invariant $\gR(\gamma,y)$ defined in \refb{ec5};}

\defn{CoulombH}{\vardef{list of charge vectors},\vardef{multiplicity vector},\newdefy }{denotes the (unevaluated) factor $H(\{\alpha_i\},\{n_i\},y)$ appearing in the formula \eqref{essp2} for 
$\Omega_{\rm tot}(\sum n_i\alpha_i,y)$  in terms of 
$\OmS(\alpha_i,y)$.}

\defn{QFact}{\vardef{n},\newdefy }{represents the (non-evaluated) $q$-deformed factorial $[n,y]!$}

\subsection{Environment variables}

\defvar{{\tt \$QuiverPerturb1}}{Sets the size of the perturbation $\epsilon_1=1/\var{\$QuiverPerturb}$ of the DSZ products in \eqref{deform1}, set to 1000 by default.}

\defvar{{\tt \$QuiverPerturb2}}{Sets the size of the perturbation $\epsilon_2=1/\var{\$DSZPerturb}$ of the DSZ products in \eqref{deform2}, set to $10^{10}$ by default.}

\defvar{{\tt \$QuiverNoLoop}}{If set to True, the quiver will be assumed to have no oriented loop, hence all $H$ factors and all $\OmS(\alpha)$ will be set to zero (unless $\alpha$ is a basis vector). Set to False by default.}

\defvar{{\tt \$QuiverTestLoop}}{If set to True, all $H$ factors and $\OmS(\alpha)$ corresponding to subquivers without loops will be set to zero (unless $\alpha$ is a basis vector).
Set to True by default.
Determining whether a subquiver has loops is time-consuming, so for large quivers it may be advisable to disable this feature. Note that 
\var{{\tt \$QuiverNoLoop}} takes precedence over this variable.}

\defvar{{\tt \$QuiverMultiplier}}{Overall scaling factor of the DSZ matrix in any evaluation of 
\var{\tt Coulombg} or \var{\tt HiggsG}. Set to 1 by default, could be a formal variable.}

\defvar{{\tt \$QuiverVerbose}}{If set to False, all consistency tests on data and corresponding error messages will be skipped. Set to True by default.}

\defvar{{\tt \$QuiverDisplayCoulombH}}{If set to True, the routine 
\fun{\tt CoulombBranchFormula} will return a list   $\{ \var{Q}, \var{R} \}$ where
 $\var{Q}$  is the Poincar\'e-Laurent   polynomial and \var{R} is a  list of replacement rules for the \var{CoulombH} factors. Set to False by default.}

\defvar{{\tt \$QuiverPrecision}}{Sets the numerical precision with which all consistency tests
are carried out. This is set to 0 by default since all data are assumed to be rational numbers. This can be set to a small real number when using real data, however the user is warned that rounding errors
tend to grow quickly.}

\defvar{{\tt \$QuiverRecursion}}{If set to 1 (default value), then the recursion relations 
based on \eqref{enewrecur}   are used for computing
\fun{\tt CoulombF};  if set to 0  the recursion relation
\eqref{edeltafs1} is used instead. } \newnote 

\defvar{{\tt \$QuiverOmSbasis}}{Set to 1 by default. If set to 0,  the routines 
\fun{\tt SimplifyOmSbasis}, \fun{\tt SimplifyOmSbasismult}, \fun{\tt OmSNoLoopToZero},  
\fun{\tt OmTNoLoopToZero},   and \fun{\tt TestNoLoop}  
are deactivated, so that 
 the assumption that basis 
vectors carry $\OmS(\ell\gamma_i)=\delta_{\ell,1}$ and quivers without loop have
$\OmS=0$ is relaxed.}

\subsection{Coulomb index}

\defn{CoulombF}{\vardef{Mat},\vardef{Cvec}}{returns the index of collinear solutions 
$F(\{\ta_1,\cdots \ta_n\}, \{\cc_1,\cdots \cc_n\})$ with 
                   DSZ products $\ta_{ij}=\var{Mat}[[i,j]]$, FI terms $\cc_i=\var{Cvec}[[i]]$ 
                   and trivial ordering.}

\defn{CoulombG}{\vardef{Mat}}{returns the index of scaling collinear solutions 
$G(\{\ha_1,\cdots \ha_n\})$ 
with  DSZ products $\ha_{ij}=\var{Mat}[[i,j]]$ and trivial ordering. The total angular momentum 
                   $\sum_{i<j} Mat[[i,j]]$ must vanish;}

\defn{CoulombIndex}{\vardef{Mat},\vardef{PMat},\vardef{Cvec},\newdefy }{evaluates the Coulomb index $\gref(\{\alpha_1,\cdots $ $\alpha_n\};$ $\{\dx_1,\cdots \dx_n\};y)$ 
                   with DSZ products $\alpha_{ij}=\var{Mat}[[i,j]]$, perturbed to \var{PMat}[[i,j]] so as to lift 
                   accidental degeneracies, possibly rescaled by an overall factor of 
                   \var{{\tt \$QuiverMultiplier}}, 
                   FI terms $\dx_i=\var{Cvec}[[i]]$, angular momentum fugacity \var{y};
                   }

\defn{CoulombFNum}{\vardef{Mat}}{computes numerically the index $F(\{\ta_1,\dots \ta_n\},\{\cc_1,\dots \cc_n\})$  with DSZ matrix $\ta_{ij}=\var{Mat}[[i,j]]$ and FI parameters $\cc_i=\var{Cvec}[[i]]$. For testing purposes only, works for up to 5 centers.}

\defn{CoulombGNum}{\vardef{Mat}}{computes numerically the scaling index $G(\ha_1,\dots \ha_n)$  with DSZ matrix $\ha_{ij}=\var{Mat}[[i,j]]$. For testing purposes only, works for up to 6 centers.}

\defn{CoulombIndexNum}{\vardef{Mat},\vardef{PMat},\vardef{Cvec},\vardef{k},\newdefy }{returns the Coulomb index $\gref(\{\alpha_1,\cdots $ $\alpha_n\};$ $\{\dx_1,\cdots \dx_n\};y)$ 
                   with DSZ products $\alpha_{ij}=\var{Mat}[[i,j]]$,  possibly rescaled by an overall factor of 
                   \var{{\tt \$QuiverMultiplier}},  
                   FI terms $\dx_i=\var{Cvec}[[i]]$, angular momentum fugacity \var{y}, by searching
                   collinear solutions numerically;    For testing purposes only, works for up to 5 centers.               }

\subsection{Coulomb branch formula}

 \defn{CoulombBranchFormula}{\vardef{Mat},\vardef{Cvec},\vardef{Nvec}}
	                                   {computes  the Dolbeault 
                   polynomial  of a quiver with DSZ products $\alpha_{ij}=\var{Mat}[[i,j]]$,
                   dimension vector $N_i=\var{Nvec}[[i]]$, 
                   FI parameters $\zeta_i=\var{Cvec}[[i]]$,
                   in terms of single-centered invariants $\OmS$.
                  This standalone routine first constructs the Poincar\'e-Laurent 
                  polynomial using \eqref{essp1},
                   evaluates the Coulomb indices $\gref$, 
                   determines the $H$ factors recursively using the minimal modification hypothesis
                   and finally replaces $y$ by $t$ in the argument of $\OmS$ to construct
                   the Dolbeault polynomial. 
If \var{{\tt \$QuiverDisplayCoulombH}} is 
                   set to True, the routine  returns a list $\{ \var{Q}, \var{R} \}$,
                                    where $\var{Q}$ is the Poincar\'e polynomial and \var{R} is a  list of replacement rules for the \var{CoulombH} factors.
                                                       For quivers without loops, the process can be sped up greatly by setting \var{\tt \$QuiverNoLoop} to True.
                   For complicated quivers
                   it is advisable to implement the Coulomb branch formula step by step, using the 
                   more elementary routines described below. 
                   }


 \defn{CoulombBranchFormulaFromH}{\vardef{Mat},\vardef{Cvec},\vardef{Nvec},\vardef{R} }
	                                   {returns the Dolbeault
                   polynomial  of a quiver with DSZ products $\alpha_{ij}=\var{Mat}[[i,j]]$,  
                   dimension vector $N_i=\var{Nvec}[[i]]$, 
FI parameters $\zeta_i=\var{Cvec}[[i]]$, using the rule 
           \var{R} to replace all \var{CoulombH} factors.}

\defn{QuiverPoincarePolynomial}{\vardef{Nvec},\newdefy }{constructs the Poincar\'e-Laurent
                   polynomial of a quiver  according to \eqref{essp1}. Coincides with
                   \fun{\tt QuiverPoincarePolynomialRat}                   for primitive dimension vector;
                 }

\defn{QuiverPoincarePolynomialRat}{\vardef{Nvec},\newdefy }{constructs the rational Poincar\'e-Laurent
                   polynomial $\bQC(\gamma;\zeta;y)$  according to \eqref{essp1};}
         
\defn{QuiverPoincarePolynomialExpand}{\vardef{Mat},\vardef{PMat},\vardef{Cvec},
\vardef{Nvec}, \vardef{Q}}{
                  evaluates the Cou-lomb 
                  indices $\gref$ and   total 
                   single-centered indices $\Omega_{\rm tot}(\alpha_i,y)$   appearing in the 
                  Poincar\'e-Laurent polynomial \var{Q} of a quiver with DSZ products
                  $\alpha_{ij}=\var{Mat}[[i,j]]$, perturbed to 
                   $\var{PMat}[[i,j]]$, 
                   dimension vector $N_i=\var{Nvec}[[i]]$, 
                 FI parameters $\zeta_i=\var{Cvec}[[i]]$,              
                  using \eqref{eindex}   and \eqref{essp2};}

\defn{CoulombHSubQuivers}{\vardef{Mat},\vardef{PMat},\vardef{Nvec},\newdefy }{computes recursively all \var{CoulombH} factors for DSZ matrix \var{Mat}, perturbed to \var{PMat},
and any dimension vector strictly less than \var{Nvec}; relies on the next two routines:}

                   \defn{ListCoulombH}{\vardef{Nvec},\vardef{Q}}{ returns
                   returns
                   a pair $\{\var{ListH},\var{ListC}\}$ where \var{ListH} is a list of \var{CoulombH}
                    factors possibly
                   appearing in the Poincar\'e-Laurent polynomial \var{Q }of a quiver with dimension vector 
                   \var{Nvec}, and \var{ListC} is the list of coefficients which multiply the monomials in 
                   $\OmS(\alpha_i,y)$ canonically associated to the $H$ factors in \var{Q}.}

\defn{SolveCoulombH}{\vardef{ListH},\vardef{ListC},
\vardef{R}}{ returns
                   a list of replacement rules for the \var{CoulombH} factors 
                   listed in \var{ListH}, by applying the minimal modification hypothesis
                   to the coefficients listed in \var{ListC}. The last argument  is 
                   a replacement rule for \var{CoulombH} factors associated to subquivers.}

\defn{MinimalModif}{\vardef{f}}{returns the symmetric Laurent polynomial which coincides 
                   with the Laurent expansion expansion of the symmetric rational function $f$ at $y=0$, up to strictly positive powers of $y$. Here symmetric means invariant under $y\to 1/y$. In practice,
\fun{\tt  MinimalModif}[\var{f}] evaluates the contour integral in \cite{Manschot:2012rx}, Eq 2.9
\be
\label{uint}
\oint \frac{{\rm d} u}{2\pi {\rm i}} \frac{(1/u-u) \, f(u)}{(1-u y)(1-u/y)}
\ee 
by deforming the contour around 0 into a sum of counters over all poles of $f(u)$ and zeros
of $(1-uy)(1-u/y)$. This trick allows to compute \eqref{uint} even if the order of the pole of $f(y)$
at $y=0$ is unknown, which happens if  \var{{\tt \$QuiverMultiplier}} is a formal variable.
                   }

\defn{SimplifyOmSbasis}{\vardef{f}}{replaces  $\OmS(\gamma,y)\to 1$ when $\gamma$ is a basis vector, unless  \var{{\tt \$QuiverOmSbasis}} is set to 0; 
}  \newnote 

\defn{SimplifyOmSbasismult}{\vardef{f}}{replaces  $\OmS(\gamma,y)\to 0$ when $\gamma$   is a non-trivial multiple of a basis vector, unless  \var{{\tt \$QuiverOmSbasis}} is set to 0;}

\defn{CoulombHNoLoopToZero}{\vardef{Mat},\vardef{f}}{sets to zero any $H$
                  factor in \var{f} corresponding to subquivers without loop, assuming DSZ products
                  $\alpha_{ij}=\var{Mat}[[i,j]]$
                  ; active only on 2-node subquivers if \var{{\tt \$QuiverTestLoop}} is set to False}

\defn{OmTNoLoopToZero}{\vardef{Mat},\vardef{f}}{sets to zero any $\Omega_{\rm tot}$
                  factor in \var{f} corresponding to subquivers without loop, assuming DSZ products
                  $\alpha_{ij}=\var{Mat}[[i,j]]$
                  ; active only on 2-node subquivers if \var{{\tt \$QuiverTestLoop}} is set to False,
                  deactivated if \var{\tt \$QuiverOmSbasis} is set to 0;
                  }

\defn{OmSNoLoopToZero}{\vardef{Mat},\vardef{f}}{sets to zero any $\OmS$  
                  factor in \var{f} corresponding to subquivers without loop, assuming DSZ products
                  $\alpha_{ij}=\var{Mat}[[i,j]]$
                  ; active only on 2-node subquivers if \var{{\tt \$QuiverTestLoop}} is set to False,  
                  deactivated if \var{\tt \$QuiverOmSbasis} is set to 0;
                  }

                   \defn{EvalCoulombH3}{\vardef{Mat},\vardef{f}}{ evaluates any 3-center $H$ factor with multiplicity vector $\{1,1,1\}$ 
               appearing   in $f$. Not used in any routine so far.}

\defn{DropFugacity}{\vardef{f}}{replaces $\OmS(\gamma,y^m,t^m)$ by $\OmS(\gamma,t^m)$ everywhere in $f$} \newnote 
                
\defn{SwapFugacity}{\vardef{f}}{replaces $\OmS(\gamma,y^m)$ with $\OmS(\gamma,y^m,t^m)$ everywhere in $f$}

\subsection{Higgs branch formula}

\defn{HiggsBranchFormula}{\vardef{Mat},\vardef{Cvec},\vardef{Nvec} }
	                                   {computes the Poincar\'e-Laurent
                   polynomial  of a quiver with DSZ products $\alpha_{ij}=\var{Mat}[[i,j]]$ (possibly              rescaled by \var{{\tt \$QuiverMultiplier}}), dimension vector $N_i=\var{Nvec}[[i]]$, 
                   FI parameters $\zeta_i=\var{Cvec}[[i]]$, using the Higgs branch formula  \eqref{eq:inversestackinv}. It is assumed, but not checked, that the quiver has no oriented 
                   loop;}

\defn{StackInvariant}{\vardef{Mat},\vardef{Cvec},\vardef{Nvec},\newdefy }{gives 
                 the stack 
                  invariant \eqref{ec5} 
                  of a quiver with DSZ matrix $\alpha_{ij}=\var{Mat}[[i,j]]$, possibly rescaled 
                  by an overall factor of 
                   \var{{\tt \$QuiverMultiplier}}, 
                  FI parameters $\zeta_i=\var{Cvec}[[i]]$,  dimension vector $N_i=\var{Nvec}[[i]]$, 		using Reineke's formula \eqref{ec5}; the answer is written in terms of unevaluated
		$q$-deformed factorials \var{QFact[n,y]};}
		
		\defn{AbelianStackInvariant}{\vardef{Mat},\vardef{Cvec},\newdefy }{gives 
                 the Abelian stack 
                  invariant \eqref{edefghiggs} 
                  of a quiver with DSZ matrix $\alpha_{ij}=\var{Mat}[[i,j]]$, possibly rescaled 
                  by an overall factor of 
                   \var{{\tt \$QuiverMultiplier}}, 
                  FI parameters $\zeta_i=\var{Cvec}[[i]]$, 		
                  using Reineke's formula \eqref{ec5}; coincides with \fun{\tt StackInvariant} 
                  with \var{Nvec}$=\{1,\dots 1\}$ except that tests of marginal or threshold stability
                  are performed (unless   \var{\tt \$QuiverVerbose} is set to False);
                  }
                   
\defn{QDeformedFactorial}{\vardef{n},\newdefy }{gives the $q$-deformed factorial $[n,y]!$}

\defn{EvalQFact}{\vardef{f}}{evaluates any \var{QFact[n,y]} appearing in \var{f}}

\subsection{Utilities}

\defn{ListAllPartitions}{\vardef{charge vector}}{returns the list of unordered 
                   partitions $\{\alpha_i\}$ of the positive integer vector $\gamma$ as a sum of positive, non-zero integer vectors $\alpha_i$; 
                   }
                   
                   \defn{ListAllPartitionsMult}{\vardef{charge vector}}{returns the list of unordered 
                   partitions $\{\alpha_i,m_i\}$ of the positive integer vector $\gamma$ as a sum of positive, non-zero integer vectors $\alpha_i$ with multiplicity $m_i$; 
                   }

\defn{ListSubQuivers}{\vardef{Nvec}}{gives a list of all dimension vectors less or equal to \var{Nvec};}

\defn{SubDSZ}{\vardef{Mat},\vardef{Cvec},\vardef{Li}}{gives the
                    DSZ matrix of the subquiver made of vectors in list \var{Li};}

 \defn{SymmetryFactor}{\vardef{Li}}{gives the symmetry factor $1/  |{\rm Aut}(\{\alpha_1, \alpha_2,\cdots, \alpha_n\}|$ for the list of charge vectors \var{Li};}

\defn{OmTRat}{\vardef{Nvec},\newdefy }{ gives the rational total invariant 
$\bar\Omega_{\rm tot}(\gamma;y)$ in terms of $\Omega_{\rm tot}(\gamma;y)$.
Coincides with the latter if $\gamma$ is primitive.}

\defn{OmTToOmS}{\vardef{f}}{expands out any $\Omega_{\rm tot}(\gamma;y)$ in $f$
 into $H$ factors and $\OmS$'s using\eqref{essp2};}
           
\defn{OmToOmb}{\vardef{f}}{expresses any $\Omega(\gamma;y)$ in $f$
 in terms of $\bar\Omega(\gamma;y)$'s;}

\defn{OmbToOm}{\vardef{f}}{expresses  any $\bar\Omega(\gamma;y)$ in $f$
 in terms of $\Omega(\gamma;y)$'s;}

\defn{HiggsGToOmb}{\vardef{Nvec},\newdefy }{Returns the (unevaluated) HN invariant
$\gR(\gamma,y)$ in terms of the rational refined indices $\Omega(\gamma;y)$
using \eqref{gtoQb};}

\defn{OmbToHiggsG}{\vardef{Nvec},\newdefy }{Returns the (unevaluated) rational refined index $\Omega(\gamma;y)$ in terms of the (unevaluated) 
stack invariants $\gR(\gamma,y)$ 
using \eqref{eq:inversestackinv};}

\defn{RandomCvec}{\vardef{Nvec}}{generates a random set of FI parameters $\zeta_i$ between -1 and 1, such that $\sum \zeta_i\, \var{Nvec}[[i]]=0$;}

\defn{UnitStepWarn}{\vardef{x}}{gives 1 for $x>0$, 0 for $x<0$, and $1/2$ if $x=0$. Produces a warning in this latter case, irrespective of the value of \var{\tt \$QuiverVerbose}. If so, the user
is advised run the 
computation again with a different random perturbation. 
}

\defn{GrassmannianPoincare}{\vardef{k},\vardef{n},\newdefy}{computes the Poincar\'e
polynomial of the Grassmannian $G(k,n)$ via Eq. (6.22) in \cite{Manschot:2012rx}.} \newnote

\defn{CyclicQuiverOmS}{\vardef{avec},\vardef{t}}{computes the single-centered index 
$\OmS(\gamma_1,\dots, \gamma_K)$ associated
to a cyclic Abelian quivers with DSZ matrix $\alpha_{i,j+1}=\var{avec}[[i,i+1]]$ via Eq
(4.29) in \cite{Manschot:2012rx}.} 

\defn{QuiverPlot}{\vardef{Mat}}{Displays the quiver with DSZ matrix \var{Mat}.} 

\subsection{Mutations} \newnote 

The following routines and environment variables were introduced in  {\tt CoulombHiggs.m} v2.0,
to allow investigation of mutations of generalized quivers \cite{Manschot:2013dua}:

\defn{MutateRight}{\vardef{Mat},\vardef{Cvec},\vardef{Nvec},\vardef{k}}{
Computes the  DSZ matrix, FI parameters and dimension vector of the quiver obtained
by applying a right-mutation with respect to the node $k$. If \var{k} is a list $\{k_i\}$,  then the
right mutations $k_i$ are applied successively, starting from the last entry in \var{k}.  No
consistency check on the FI parameters is performed.}

\defn{MutateLeft}{\vardef{Mat},\vardef{Cvec},\vardef{Nvec},\vardef{k}}{
Computes the  DSZ matrix, FI parameters and dimension vector of the quiver obtained
by applying a left-mutation with respect to the node $k$. If \var{k} is a list $\{k_i\}$,  then the
right mutations $k_i$ are applied successively, starting from the last entry in \var{k}. No
consistency check on the FI parameters is performed.} 

\defn{OmStoOmS2}{\vardef{f}}{replaces $\var{\tt OmS[gam,y,t]}$ by $\var{\tt OmS2[gam,y,t]}$
anywhere in $\var{f}$. This is useful for distinguishing the single-centered invariants of the mutated
quiver from those of the original one.}

\defn{MutateRightOmS}{\vardef{Mat},\vardef{k},\vardef{f}}{expresses the single-centered
invariants $\var{\tt OmS[gam,y,t]}$ of the original quiver with DSZ matrix \var{Mat} in terms of the single-centered
invariants $\var{\tt OmS2[gam,y,t]}$ of the quiver obtained by right-mutation with respect to node $k$,
using Eq. 1.13 in \cite{Manschot:2013dua}.}

\defn{MutateLeftOmS}{\vardef{Mat},\vardef{k},\vardef{f}}{expresses the single-centered
invariants $\var{\tt OmS[gam,y,t]}$ of the original quiver with DSZ matrix \var{Mat} in terms of the single-centered
invariants $\var{\tt OmS2[gam,y,t]}$ of the quiver obtained by left-mutation with respect to node $k$,
using Eq. 1.13 in \cite{Manschot:2013dua}.}

\defn{MutateRightOmS2}{\vardef{Mat},\vardef{k},\vardef{f}}{expresses the single-centered
invariants $\var{\tt OmS2[gam,y,t]}$ a quiver with DSZ matrix \var{Mat} in terms of the single-centered
invariants $\var{\tt OmS[gam,y,t]}$ of the quiver obtained by right-mutation with respect to node $k$.
Identical to \fun{\tt MutateRightOmS}, except for swapping $\var{\tt OmS[gam,y,t]}$ and $\var{\tt OmS2[gam,y,t]}$.}

\defn{MutateLeftOmS2}{\vardef{Mat},\vardef{k},\vardef{f}}{expresses the single-centered  
invariants $\var{\tt OmS2[gam,y,t]}$ a quiver with DSZ matrix \var{Mat} in terms of the single-centered
invariants $\var{\tt OmS[gam,y,t]}$ of the quiver obtained by right-mutation with respect to node $k$.
Identical to \fun{\tt MutateLeftOmS}, except for swapping $\var{\tt OmS[gam,y,t]}$ and $\var{\tt OmS2[gam,y,t]}$.}

\defn{DropOmSNeg}{\vardef{f}}{equates to 0 any $\OmS(\gamma,y,t)$ where the dimension
vector associated to $\gamma$ has negative components.}

\defvar{{\tt \$QuiverMutationMult}}{Equal to 1 by default. Set to $M$, defined in Eq. (1.8) of
  \cite{Manschot:2013dua} when dealing with generalized quivers.}

\end{document}